\documentclass{elsart}

\usepackage{float}
\usepackage{amssymb}
\usepackage{amsfonts}
\usepackage{amsmath}
\usepackage{relsize}
\usepackage{rotating}
\usepackage{graphicx}
\usepackage{caption}
\usepackage{subcaption}
\usepackage{bm}
\usepackage{color}
\usepackage{multirow}
\usepackage{tikz}
\usetikzlibrary{decorations.markings}
\usetikzlibrary{shapes,arrows}
\usetikzlibrary{shapes.geometric}
\usetikzlibrary{fit}
\usetikzlibrary{backgrounds}
\usepackage{setspace}
\usepackage{algorithm}
\usepackage{algpseudocode}
\usepackage[utf8]{inputenc}
\usepackage[english]{babel}
\usepackage{cite}
\renewcommand{\figurename}{\textbf{Fig.}}
\renewcommand{\tablename}{\textbf{Table}}
\begin{document}
\setcounter{page}{1}
\allowdisplaybreaks[4]

\begin{frontmatter}

\title{Multiscale Extended Finite Element Method for Deformable Fractured Porous Media}

\author[TUD]{Fanxiang Xu\corauthref{cor1}}
\ead{f.xu-4@tudelft.nl}
\author[TUD2]{Hadi Hajibeygi}
\ead{H.Hajibeygi@tudelft.nl}
\author[TUD]{Lambertus J. Sluys}
\ead{L.J.Sluijs@tudelft.nl}

\address[TUD]{Materials, Mechanics, Management and Design, Faculty of Civil Engineering and Geosciences, Delft University of Technology, P.O. Box 5048, 2600 GA Delft, The Netherlands}
\address[TUD2]{Department of Geoscience and Engineering, Faculty of Civil Engineering and Geosciences, Delft University of Technology, P.O. Box 5048, 2600 GA Delft, The Netherlands}

\corauth[cor1]{Corresponding author.}

\begin{abstract}
Deformable fractured porous media appear in many geoscience applications. While the extended finite element (XFEM) has been successfully developed within the computational mechanics community for accurate modeling of the deformation, its application in natural geoscientific applications is not straightforward. This is mainly due to the fact that subsurface formations are heterogeneous and span large length scales with many fractures at different scales. In this work, we propose a novel multiscale formulation for XFEM, based on locally computed basis functions. The local multiscale basis functions capture the heterogeneity and discontinuities introduced by fractures. Local boundary conditions are set to follow a reduced-dimensional system, in order to preserve the accuracy of the basis functions. Using these multiscale bases, a multiscale coarse-scale system is then governed algebraically and solved, in which no enrichment due to the fractures exist. Such formulation allows for significant computational cost reduction, at the same time, it preserves the accuracy of the discrete displacement vector space. The coarse-scale solution is finally interpolated back to the fine scale system, using the same multiscale basis functions. The proposed multiscale XFEM (MS-XFEM) is also integrated within a two-stage algebraic iterative solver, through which error reduction to any desired level can be achieved. Several proof-of-concept numerical tests are presented to assess the performance of the developed method. It is shown that the MS-XFEM is accurate, when compared with the fine-scale reference XFEM solutions. At the same time, it is significantly more efficient than the XFEM on fine-scale resolution. As such, it develops the first scalable XFEM method for large-scale heavily fractured porous media.

\end{abstract}

\end{frontmatter}

\section{Introduction}%

Subsurface geological formations are often highly heterogeneous and heavily fractured at multiple scales. Heterogeneity of the deformation properties (e.g. elasticity coefficients) can be of several orders of magnitude which occurs at fine scale (cm) resolution. The reservoirs also span large scales, in the order of kms. Numerical simulation of mechanical deformation for such complex systems is necessary to optimise the geo-engineering operations \cite{Zoback_95,Ernest1}, and assess their safety and manage the associated risks (e.g. fracture propagation, fault slip and induced seismicity). Though being crucially important, simulation of these systems are beyond the scope of classical numerical schemes.

Presence of highly heterogeneous coefficients with high resolution within large-scale domains has been systematically addressed in the computational geoscience community through the development of multiscale finite element and finite volume methods \cite{Multiscale1_Hadi,Nicolai_MSFEM,Sokolova2019,Jenny2003}. Recent developments also include mechanical deformation coupled with fluid pore pressure dynamics \cite{Deb2017,Ren2016,Castelletto2016,Fumagalli2014,Giovanardi2017}. 
In presence of many fractures, however, the complexity of the computational model increases significantly. As such, development of a robust multiscale strategy for deformation of heavily fractured porous media, which also allows for convergent systematic error reduction\cite{Multiscale2_Hadi,YWang2014,CHUNG201454}, is of high interest in the geoscience community.

The presence of fractures within the computational domain can be included explicitly by two approaches of (1) unstructured grid and (2) immersed or embedded methods. 
\newline The unstructured grid approach \cite{Rashid1998,Bittencourt1996,Cook1995} generates a discrete computational domain in which fractures are always at the interfaces of elements. This allows for convenient treatment of their effect, however, at the cost of complex meshing. The complex mesh generation for three-dimensional (3D) large scale domains with many fractures is challenging, specially when fractures dynamically extend their geometries. On the other hand, the immersed or embedded approach allows for independent grids for matrix block and fractures, by introducing enrichment of the discrete connectivity (for flow) and shape functions (for mechanics)\cite{Wells2001,Tene2017,Khoei2014,Efendiev2014,Wu2015}. These enriched formulations are aimed at representing discontinuities within the overlapping matrix cell, without any adjustment nor refinement of the grid\cite{Belytschko1994}. The enrichment strategy for modeling deformation using finite-element schemes in presence of fractured media are referred to as 'extended finite element (XFEM)' methods. 

XFEM enriches the partition of unity (PoU) 
\cite{MELENK1996} by introducing additional degrees of freedom (DOF) at the existing element nodes. There exists sets of enriched functions to capture the jump discontinuity in the displacement field, when the fracture element cuts through the entire cell, and the tip when a fracture ends within the domain of an element (i.e. its tip is inside an element). \cite{Moes1999,Belytschko_XFEM,Aragon2017,FPMeer2009,GNwells}. For these jump and tip scenarios, additional shape functions are introduced which are multiplied by the original shape functions and supplement the discrete displacement approximation space. 

When it comes to geoscience applications, the XFEM is not an attractive method, due to its excessive additional degrees of freedom to capture the many fractures. As such one has to develop a scalable approach, in order to allow for accurate yet efficient application of XFEM to simulate deformation in geological formations. 

This paper develops a multiscale XFEM (referred to as MS-XFEM) which offers a scalable efficient strategy to model large-scale fractured systems. MS-XFEM imposes a coarse mesh on the given fine-scale mesh. The main novel idea behind MS-XFEM is to use XFEM to computationally solve for local coarse-scale (multiscale) basis functions. These basis functions capture the fractures and coefficient heterogeneity within each coarse element. The solving strategy of these local coarse-scale basis functions can be either geometric or algebraic \cite{HosseiniMehr2020,HosseiniMehr2018,YWang2014}. We prefer algebraic construction, as it allows for black-box integration of the method within any existing XFEM simulator. Once the basis functions are solved, they will be clustered in the matrix of Prolongation (P), which maps the coarse-scale solution to the fine-scale one. Note that there will be no additional multiscale basis functions due to jump or tips, and that only 4 multiscale basis functions per element exist for 2D structured grids (8 in 3D) in each direction (x, y, and z).  

The fine-scale XFEM system is then mapped to the coarse grid by using the Restriction (R) operator, which is defined based on the FEM, as the transpose of the prolongation operator. The approximate fine-scale solution is finally obtained after mapping the coarse-scale solution to the fine scale, by using the prolongation operator. 

The approximate solution of MS-XFEM can be found acceptable for many applications, however error control and reduction to any desired level is necessary to preserve its applicability for challenging cases. As such, the MS-XFEM is integrated within the two-stage iterative solver in which the MS-XFEM is paired with an efficient iterative smoother (here ILU(0)) to reduce the error\cite{Chow1997,Zhou2012}. One can also use the Krylov subspace methods (e.g. GMRES) to enhance the convergence, which stays outside the scope of this paper.  

Several proof-of-concept numerical tests are presented to assess the accuracy of the presented MS-XFEM without and with iterative improvements. The test cases include large deformations which may not be realistic in geoscience applications, but important to be studied in order to quantify the errors in large deformation scenarios. From these results it becomes clear that the MS-XFEM, despite using no enriched basis functions at coarse scale, presents an efficient and accurate formulation to study deformation of fractured geological media.  

The structure of this paper is set as the following. Next, the governing equations and the fine scale XFEM method are introduced. Then, the MS-XFEM method is presented in detail, with emphasis on the construction of local multiscale basis function and the approximate fine scale solution. Then, different numerical test cases are presented. Finally, concluding remarks are discussed.

\section{Governing Equations and Fine-scale XFEM System}
Consider the domain $\Omega$ bounded by $\Gamma$ as shown in figure \ref{comp_domain}. Prescribed displacements or Dirichlet boundary condition are imposed on $\Gamma_u$, while tractions are imposed on $\Gamma_t$. The crack surface $\Gamma_c$ (lines in 2-D and surfaces in 3-D) is assumed to be traction-free. 
\begin{center}
	\centering
	\includegraphics[trim={0cm 0cm 0cm 0cm}, clip, width=0.6\textwidth]{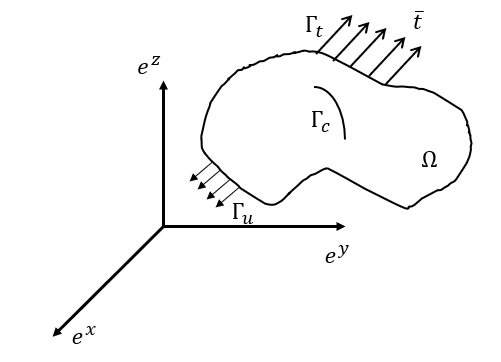}
	\captionof{figure} {An illustration of fractured domain setup}\label{comp_domain}
\end{center}

The momentum balance equations and boundary conditions read
\begin{align} 
\nabla \cdot \sigma + f =0 \thickspace \ \ \ in \ \Omega \label{gov_1}\\ 
\sigma\cdot \overrightarrow{n} =\bar{t} \thickspace \ \ \ on \ \Gamma_t \\
\sigma\cdot \overrightarrow{n} =0  \thickspace \ \ \ on \ \Gamma_c \\
u=\bar{u}  \thickspace \ \ \ on \ \Gamma_u,
\end{align}
where $\sigma$ is the stress tensor and $u$ is the displacement field over the whole domain. $\overrightarrow{n}$ is the normal vector pointing outside the domain \cite{White_2020,TEREKHOV2020112357}.\\
The constitutive law with linear elasticity assumption reads
\begin{equation} \label{elastic}
\sigma=C: \varepsilon=C: \nabla^{s} u
\end{equation}
where, $\nabla^{s}$ denotes the symmetrical operator and $C$ is the property tensor defined as
\begin{gather}
\nonumber
C=
\begin{bmatrix} 
\lambda+2\mu & \mu & 0 \\  
\mu & \lambda+2\mu & 0 \\ 
0 & 0 & \mu 
\end{bmatrix},
\end{gather}
with $\lambda$ and $\mu$ denoting the Lame's parameters \cite{wang2017,GASPAR2003487}.

The strain tensor $\varepsilon$ is expressed as
\begin{equation}
\label{strain_h}
\varepsilon = \nabla^{s} u=\frac{1}{2} (\nabla u+\nabla^{T} u)
\end{equation}
where, $\nabla$ denotes the gradient operator.

Substituting Eqs. \eqref{elastic} and \eqref{strain_h} in the governing equation \eqref{gov_1} results in a 2nd order Partial Differential Equation (PDE) for displacement field $u$
\begin{equation}
\label{displacementEq}
\nabla \cdot (C : \nabla^s u) + f = 0.
\end{equation}

Equation \eqref{displacementEq} is then solved for computational domains with cracks (representing faults and fractures). This is done by the extended finite element (XFEM) method, which is briefly revisited in the next section.

\subsection{Extended Finite Element Method (XFEM)}
The FEM with smooth shape functions $N_i$ provides an approximate numerical solution to Eq. \eqref{displacementEq} for displacement unknowns, i.e.,
\begin{equation}\label{fem}
u=\sum_{i\in I} u_i N_i.
\end{equation}
This formula can be used for computational domains without discontinuity. The FEM approximation is insufficient to capture discontinuities imposed by the existence of the fractures and faults. As such, the XFEM method introduces two sets of enrichment to the original FEM in order to allow it to capture the discontinuities without adapting the grid. These enrichment sets are associated with the body and tip of the fractures and faults. The body is enriched by jump functions, and the tip by tip enrichment functions \cite{Moes1999}. Below brief descriptions of these two enrichment functions are provided.
\subsubsection{Jump enrichment}
The jump enrichment represents the discontinuity involved in the displacement field across the fracture and fault main body. The jump enrichment is often chosen as the step or Heaviside function, which can be expressed as
\[
H(x) =
\begin{cases}
1 & \text{on {$\Omega^+$}} \\
-1 & \text{on {$\Omega^-$}}\\
\end{cases}
\].

Note that $\Omega^+$ and $\Omega^-$ zones are determined based on the normal vector pointing out of the fracture curve. For line fractures, the direction can be any side, as long as all discrete elements use the same + and - sides for a fracture. 

\subsubsection{Tip enrichment}
The tip enrichment represents the discontinuity of the displacement field near the fracture tip. This type of enrichment function, denoted by $F_l$, is based on the auxiliary displacement field near the fracture tip and contains four functions, i.e.,
\begin{equation}
{F_l(r,\theta)}=\{
{\sqrt{r} sin(\frac{\theta}{2})}, 
{\sqrt{r} cos(\frac{\theta}{2})}, 
{\sqrt{r} sin(\frac{\theta}{2}) sin(\theta)}, 
{\sqrt{r} cos(\frac{\theta}{2}) sin(\frac{\theta}{2})}\}.
\end{equation}
These four functions around the fracture tip inside the element are plotted in Figure \ref{tip_func}. The red segment, shown on the base of the plots, represents the fracture which ends in the element. Note that only the ${\sqrt{r} sin(\frac{\theta}{2})}$ contains a discontinuity around the fracture tip, while other functions are smooth.\\

\begin{center}
	\centering
	\includegraphics[trim={1cm 0cm 1cm 0cm}, clip, width=0.8\textwidth]{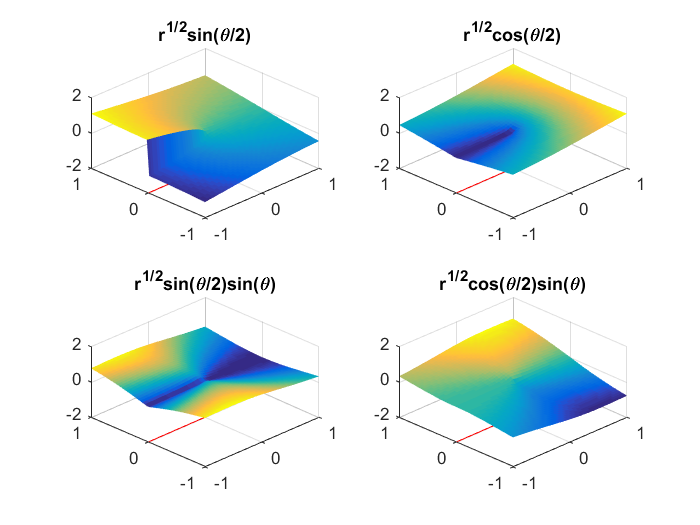}
	\captionof{figure} {Four types of tip enrichment functions inside the element. The red segment represents the crack with its tip located at center point (0,0). The discontinuity can be seen clearly in the top left function, ${\sqrt{r} sin(\frac{\theta}{2})}$.}
	\label{tip_func}
\end{center}
\subsubsection{Enrichment mechanism}
To decide whether the node is enriched or not, the node location related to the fracture is the key factor. The sketch of the enrichment mechanism is shown in Figure \ref{enrich_illustrate}. More precisely, in this figure, the tip and jump enriched nodes are highlighted in red and black, respectively. 

\begin{center}
	\centering
	\includegraphics[trim={4cm 16.5cm 5cm 2.9cm}, clip, width=0.4\textwidth]{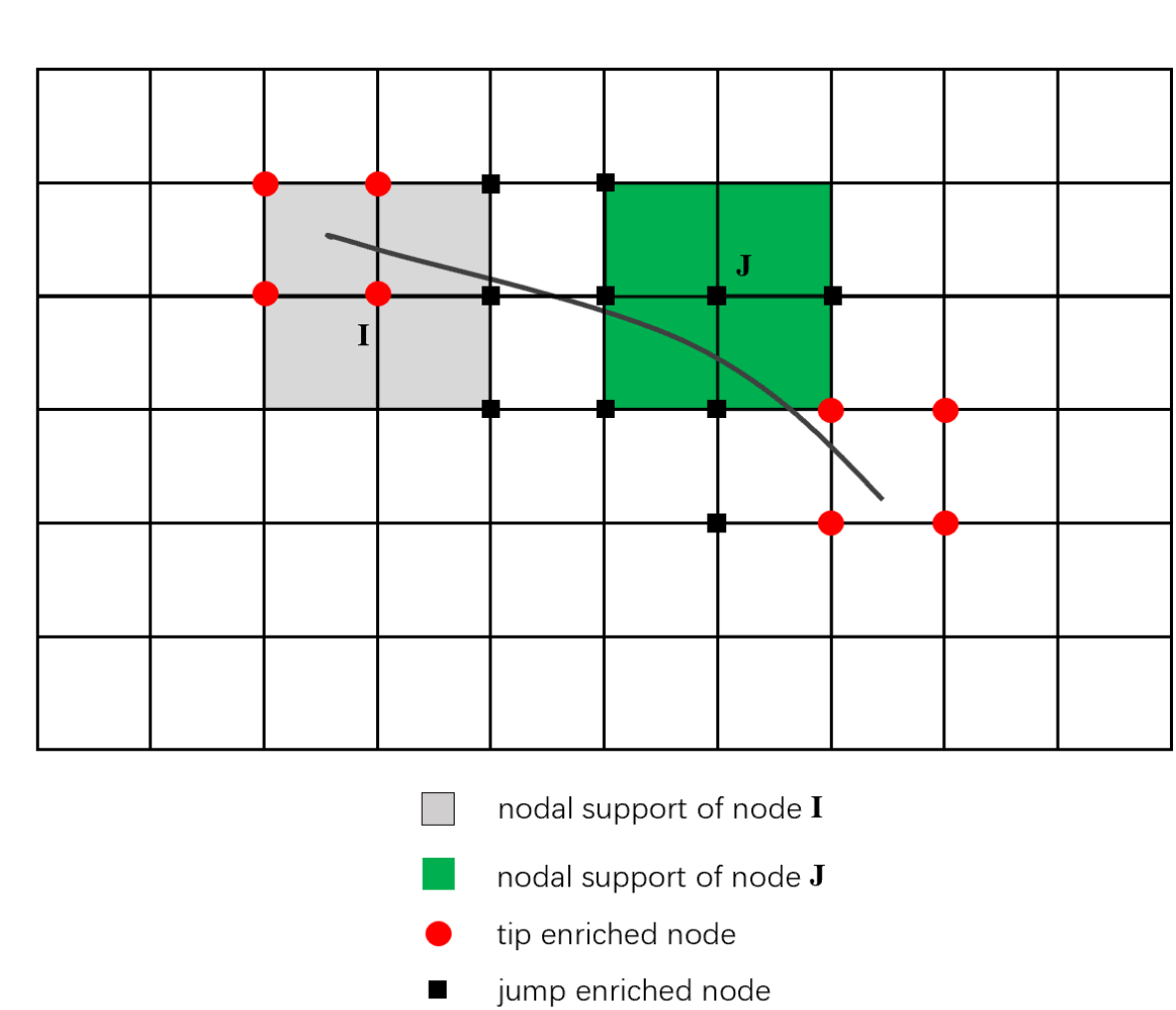}
	\captionof{figure} {Enrichment mechanism: node I and J will be enriched using tip and jump functions.}
	\label{enrich_illustrate}
\end{center}

\subsection{XFEM linear system}
The XFEM approximates the continuum displacement field $u$ at fine-scale mesh resolution $h$ by $u^h$ which is defined as
\begin{equation} \label{XFEM_formula}
u \approx u^h=\sum_{i\in \Omega^h} u_i N_i + \sum_{j\in J} a_j N_j H(x) + \sum_{k\in K}N_k,
\ \big[\sum_{l=1}^{4} F_l(x) \ b_k^{l} \big]
\end{equation}
where $N$, $H$ and $F_l$ represent, respectively, the classical FEM shape functions, the Heaviside function and  the tip enrichment functions. The fine-scale mesh has $\Omega^h$ nodes. Moreover, $u$ denotes the standard degrees of freedom (DOFs) associated to the classical finite element method. $a$ denotes the extra DOFs associated to the jump enriched node. For the jump enriched nodes, in the 2D domains, each node would contain 2 extra DOFs. Furthermore, $b$ indicates the extra DOFs associated to the tip enrichment, which adds four extra DOFs per direction (8 in total in a 2D domain) for each tip inside an element. 

The first term in the right-hand-side (RHS) of Eq. \eqref{XFEM_formula} is the contribution of the classical finite element method. This term captures the smooth deformation, using classical shape functions. The second term, however, represents the contribution of the jump enrichment. Note that the jump enrichment is modeled by the weighted Heaviside functions, with weights being the classical shape functions. There will be as many jump enrichment functions as the number of fractures inside an element. Finally, the third term in the RHS is the contribution of the fracture tips. Note that if several fracture tips end up in an element, there will be 4 additional DOFs per tip per direction in that element. 

The resulting linear system entails the nodal displacement unknowns $u$, as well as the jump level $a$ and tip weight $b$ per fracture (and fault). The augmented XFEM linear system $K^h d^h = f^h$, therefore, reads
\begin{gather}\label{XFEM_system}
\underbrace{\begin{bmatrix} 
\overline{K}_{uu} & \overline{K}_{ua} & \overline{K}_{ub} \\  \overline{K}_{au} & \overline{K}_{aa} & \overline{K}_{ab} \\ \overline{K}_{bu} & \overline{K}_{ba} & \overline{K}_{bb}
\end{bmatrix}}_{K^h} 
\underbrace{
\begin{bmatrix}
\overline{u} \\
\overline{a} \\
\overline{b}
\end{bmatrix}}_{d^h}
=
\underbrace{
\begin{bmatrix}
\overline{f}_u \\
\overline{f}_a \\
\overline{f}_b
\end{bmatrix}}_{f^h}.
\end{gather}
Compared to the classical FEM, there exist several additional blocks involved in the stiffness matrix, due to the existence of the discontinuities. The advantage of XFEM is that it does not rely on complex mesh geometry, instead, it allows fractures to overlap with the matrix elements. On the other hand, for geoscientific fractured systems, the additional DOFs due to the enrichment procedure results in excessive computational costs. This imposes a significant challenge for the XFEM application in geoscience applications. In this paper, we develop a scalable multiscale procedure which constructs a coarse-scale system based on locally supported basis functions. The method is described in the next section.

\section{Multiscale Extended Finite Element Method (MS-XFEM)}
A multiscale formulation provides an approximate solution $u'^h$ to the fine-scale XFEM deformation $u^h$ through
\begin{equation} \label{MS-XFEM_formula}
u^h \approx u'^h = \sum_{i\in \Omega^H} N^H_i u^H_i,
\end{equation}
where $N^H_i$ are the coarse-scale (multiscale) basis functions and $u^H_i$ are the coarse-scale nodal displacements at coarse mesh $\Omega^H$. Note that this multiscale formulation does not include any enrichment functions. Instead, all enrichment functions are incorporated in the construction of accurate coarse-scale basis functions $N^H$. This allows for significant computational complexity reduction, and makes the entire formulation attractive for field-scale geoscientific applications.

Next,  construction of the coarse-scale system and the basis functions will be presented.

\subsection{Coarse scale linear system}
MS-XFEM solves the linear deformation system on a coarse mesh, imposed on a given fine-scale mesh, as shown in figure \ref{MS_XFEM_mesh}. The coarsening ratio is defined as the ratio between the coarse mesh size and fine-scale mesh size. 

\begin{center}
	\centering
	\includegraphics[trim={5cm 17cm 5cm 2.9cm}, clip, width=0.5\textwidth]{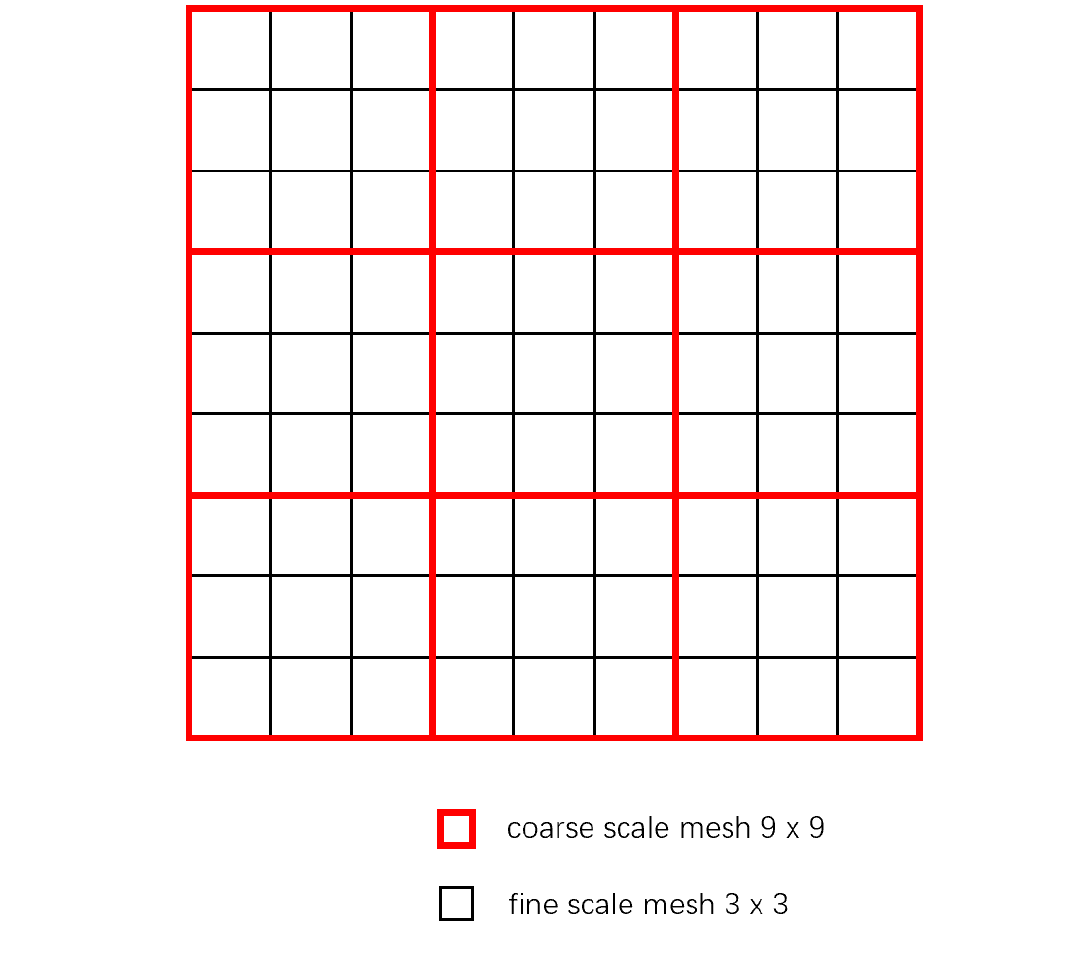}
	\captionof{figure}{Illustration of the multiscale mesh imposed on the given fine-scale mesh, with the coarsening ratio of $3\times3$.}
	\label{MS_XFEM_mesh}
\end{center}

The multiscale formula \eqref{MS-XFEM_formula} can be algebraically expressed as
\begin{equation}\label{approximate_MSXFEM}
u^h \approx u'^h = \mathbf{P} \ d^H,
\end{equation}
where $\mathbf{P}$ is the matrix of basis functions (i.e., prolongation operator) and  $d^H$ is the coarse-scale deformation vector for $u^H$ unknowns.  Algebraic formulation allows for convenient implementation of the proposed MS-XFEM, and its integration as a black-box tool for any given classical XFEM solver. Therefore, the remainder of the article will be devoted to the formulation.

The coarse-scale solution $d^H$ needs to be found by solving a coarse-scale system. To construct the coarse-scale system and solve for $d^H$, one has to restrict (map) the fine-scale linear system($K^h d^h=f^h$) to the coarse-scale, i.e.,
\begin{equation}
 \underbrace{(\mathbf{R} \ K^h \ \mathbf{P})}_{K^H} \ d^H = \mathbf{R} \ f^h.
\end{equation}
Here, $\mathbf{R}$ is the restriction operator with the size of $\Omega^H \times \Omega^{h+j+t}$, where $\Omega^{h+j+t}$ is the size of the fine-scale enriched XFEM system including jump and tip enrichment. Prolongation operator $\mathbf{P}$ has the dimension of $\Omega^{h+j+t} \times \Omega^H$. This results in the coarse-scale system matrix $K^H$ size of $\Omega^H \times \Omega^H$.

The finite-element-based restriction function is introduced as the transpose of the prolongation matrix, i.e.,
\begin{equation}
   \mathbf{R} = \mathbf{P}^T.
\end{equation}
Therefore, the coarse-scale matrix $K^H$ is symmetric-positive-definite (SPD), if $K^h$ is SPD.
Once the coarse-scale system is solved on $\Omega^H$ space for $d^H$, one can find the approximate fine-scale solution using Eq. \eqref{approximate_MSXFEM}. Overall, the multiscale procedure can be summarised as finding an approximate solution $d'^h$ according to
\begin{equation}\label{multiscale_algebraic_finalexp}
 d^h \approx d'^h = \mathbf{P} d^H = \mathbf{P} (\mathbf{R} \ K^h \ \mathbf{P})^{-1} \mathbf{R} \ f^h.
\end{equation}

Next, the prolongation operator $\mathbf{P}$, i.e., the basis functions are explained in detail. Once $\mathbf{P}$ is known, all terms in Eq. \eqref{multiscale_algebraic_finalexp} are defined.
\subsection{Construction of multiscale basis functions}
To obtain the basis functions, the governing equation \eqref{displacementEq} without any source term using XFEM, i.e., \eqref{XFEM_system} needs to be solved in each coarse element $\omega^H$. This can be expressed as solving
\begin{equation}\label{basis_c_1}
\nabla\cdot(C:(\nabla^{S} N_i^{H}))=0\thickspace \ \ \ \text{in} \ \ \Omega^H,
\end{equation}
subject to local boundary conditions. Here, we develop a reduced-dimensional equilibrium equation to solve for the boundary cells \cite{Nicola_2019,Sokolova2019}, i.e.,
\begin{equation}\label{basis_c_2}
\nabla_{\parallel}\cdot(C_r:(\nabla_{\parallel}^{S} N_i^{H}))=0\thickspace  \ \ \ \text{on} \ \ \Gamma^H.
\end{equation}
Here, $\Gamma^H$ denotes the boundary cells of the coarse element $\Omega^H$. In addition, $\nabla_{\parallel}^{S}$ denotes the reduced dimensional divergence and symmetrical gradient operators, which act parallel to the direction of the local domain boundary. 
For 2D geometries, the reduced-dimensional boundary condition represents the 1D (rod) deformation model along the coarse element edges. Note that the local basis functions involve transverse equilibrium, i.e., therefore the prolongation matrix $\mathbf{P}$ reads
\begin{gather}
\mathbf{P} =
\begin{bmatrix} 
{P}_{xx} & {P}_{xy} \\  
P_{yx} & P_{yy}  
\end{bmatrix}.
\end{gather}

\begin{figure}[H]
	\centering
	\begin{subfigure}{0.49\textwidth}
		\centering
		\includegraphics[trim={5cm 16.5cm 5cm 2.9cm}, clip, width=\textwidth]{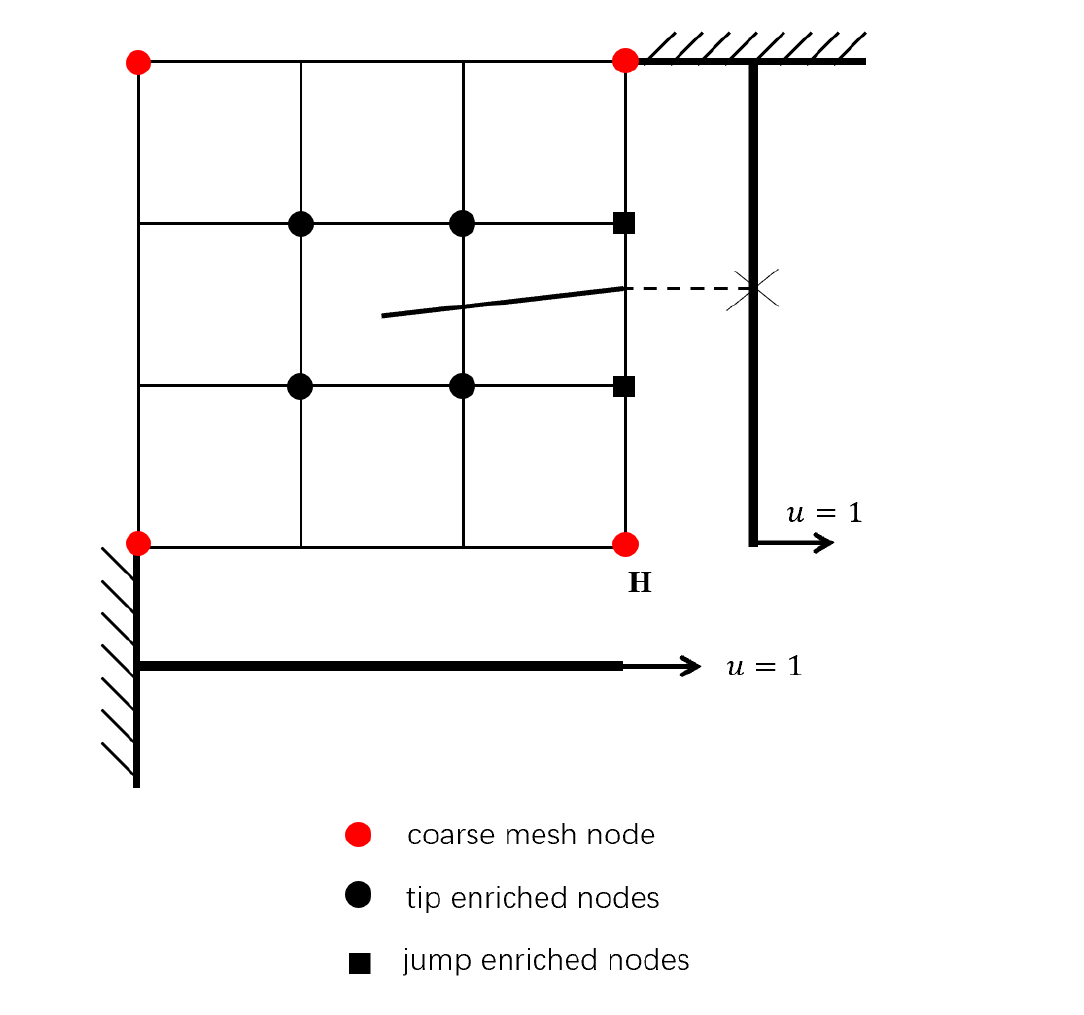}
		\caption{}
	\end{subfigure}
	\begin{subfigure}{0.49\textwidth}
	\centering
	\includegraphics[trim={5cm 16.5cm 5cm 2.9cm}, clip, width=\textwidth]{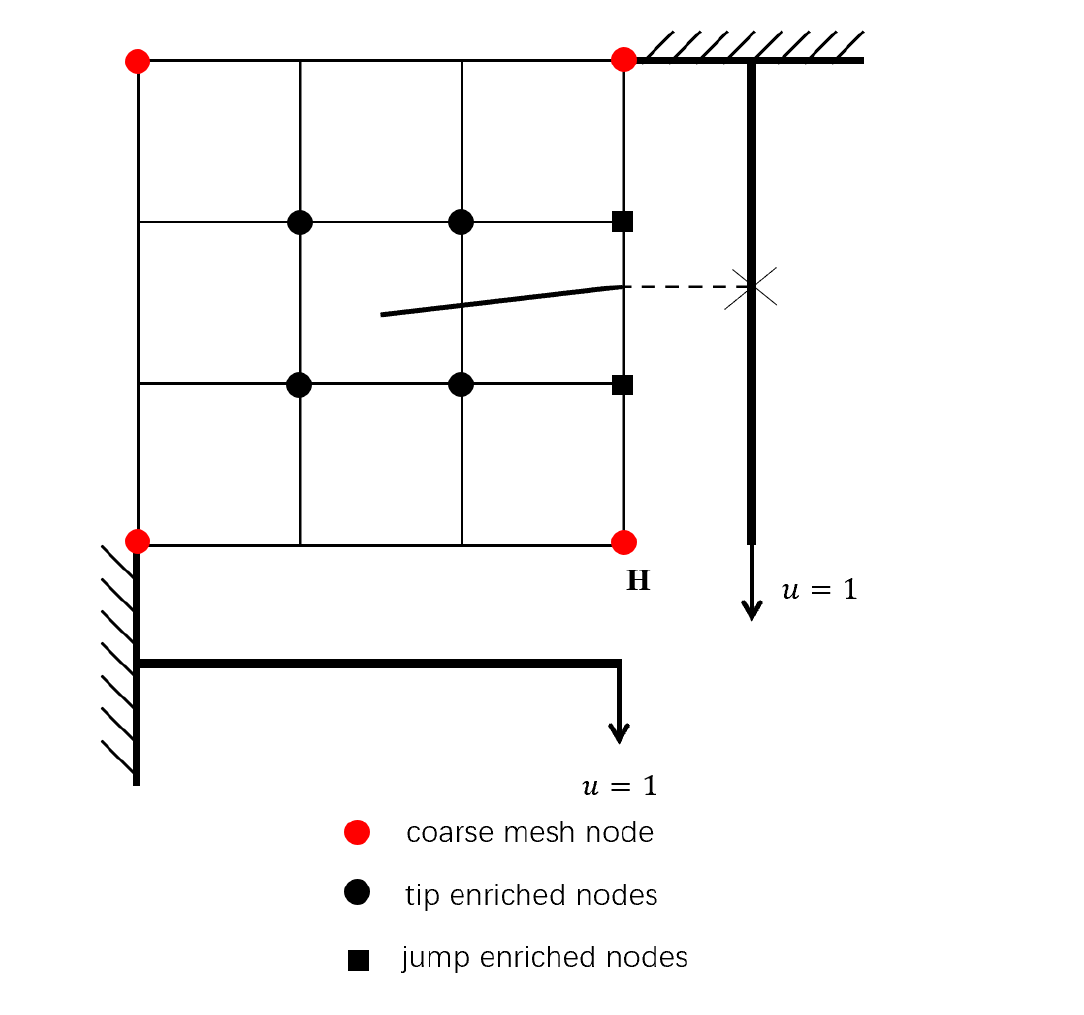}
	\caption{}
	\end{subfigure}
	\caption {Illustration of the multiscale local basis functions, constructed using XFEM  for the node $H$ in x direction (a) and y direction (b).}\label{basis_illustrate_1}
\end{figure}
Figure \ref{basis_illustrate_1} shows an example of a local system to be solved for basis functions belonging to the highlighted node $H$ in x and y directions.  Note that the Dirichlet value of 1 is set at $H$ for each directional basis functions, while all other 3 coarse mesh nodes are set to 0.\\

Note that, as shown in Fig. \ref{basis_illustrate_1}, the boundary problem is solved for both edges which have the node $H$ at one of their end values. More precisely, e.g., to find the basis function in x-direction for node $H$, we set the value of $u_x (H) = 1$ at the location $H$. This causes extension of the horizontal boundary cells and bending of the vertical boundary. \\

Once the boundary values are found, the internal cells are solved subjected to Dirichlet values for the boundary cells. An illustration of a basis function obtained using this algorithm is presented in figure \ref{basis_1}.

\begin{center}\label{basis_1}
	\centering
	\includegraphics[trim={4cm 17cm 4cm 3cm}, clip, width=0.7\textwidth]{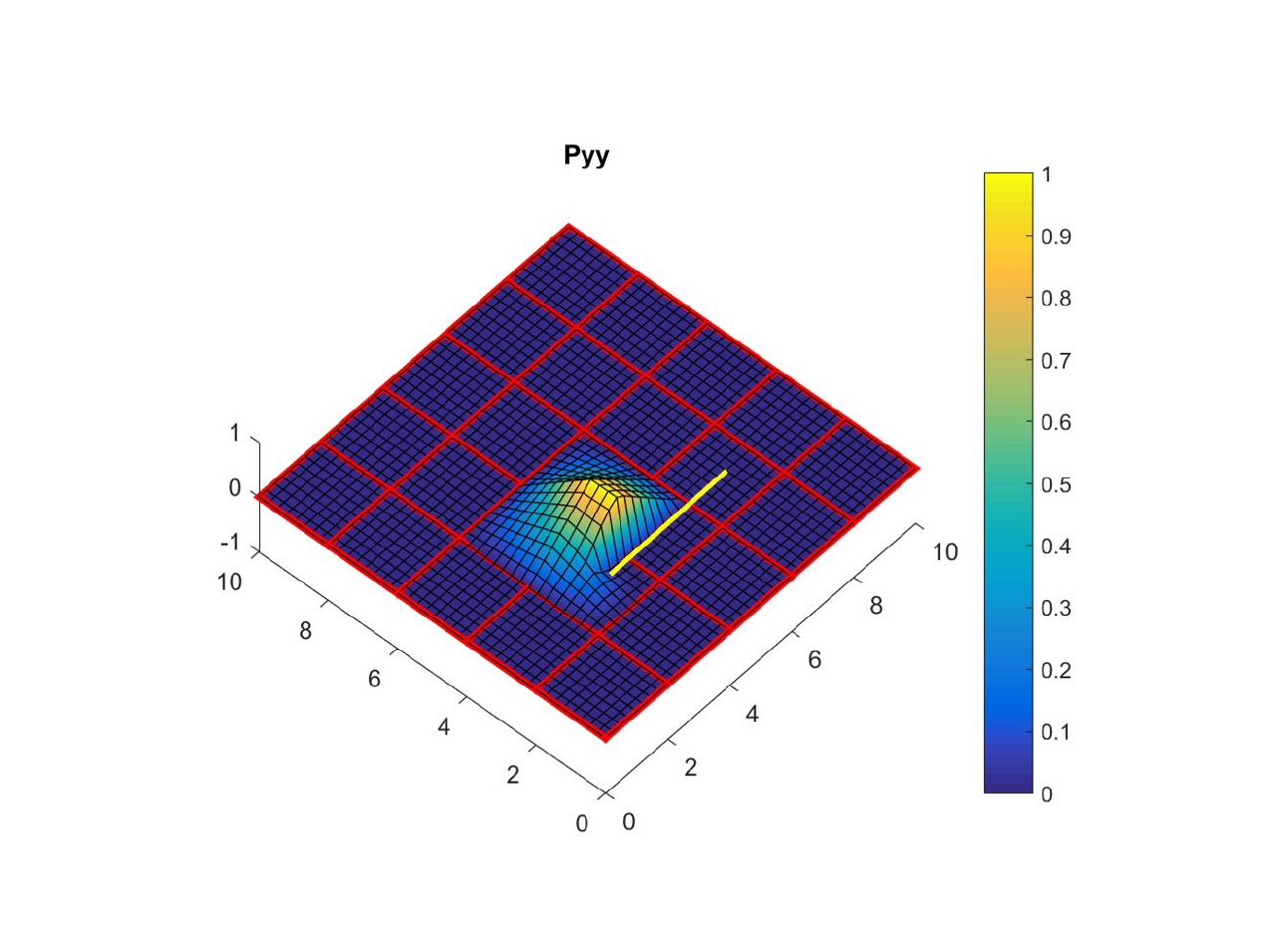}
	\captionof{figure} {Illustration of a basis function that captures the discontinuity of a fracture. Yellow segment represents the fracture.}\label{basis_1}
\end{center}

Note that the illustrated basis function captures the fractures, because of the XFEM enrichment procedure. 
The basis function $N^H_i$ will be stored in the column $i$ of the prolongation operator $\mathbf{P}$. Once all basis functions are found, the operator $\mathbf{P}$ is also known and one can proceed with the multiscale procedure as explained before.

Next, we explain how the basis functions can be algebraically computed based on the given XFEM fine-scale system. This crucial step allows for convenient integration of our multiscale method into a given XFEM simulator.

\subsection{Algebraic construction of multiscale basis functions}
The basis function formulation \eqref{basis_c_1} subjected to the local boundary condition \eqref{basis_c_2} can be constructed and solved purely algebraically. This is important, since it allows for convenient integration of the devised multiscale method into any existing XFEM simulator.

Consider the coarse cell (local domain) as shown in figure \ref{localdomain_ill}. The cells are split into 3 categories of internal, edge and vertex (node), depending on their locations \cite{YWang2014}.

\begin{center}
	\centering
	\includegraphics[trim={5cm 17cm 5cm 4cm}, clip, width=0.6\textwidth]{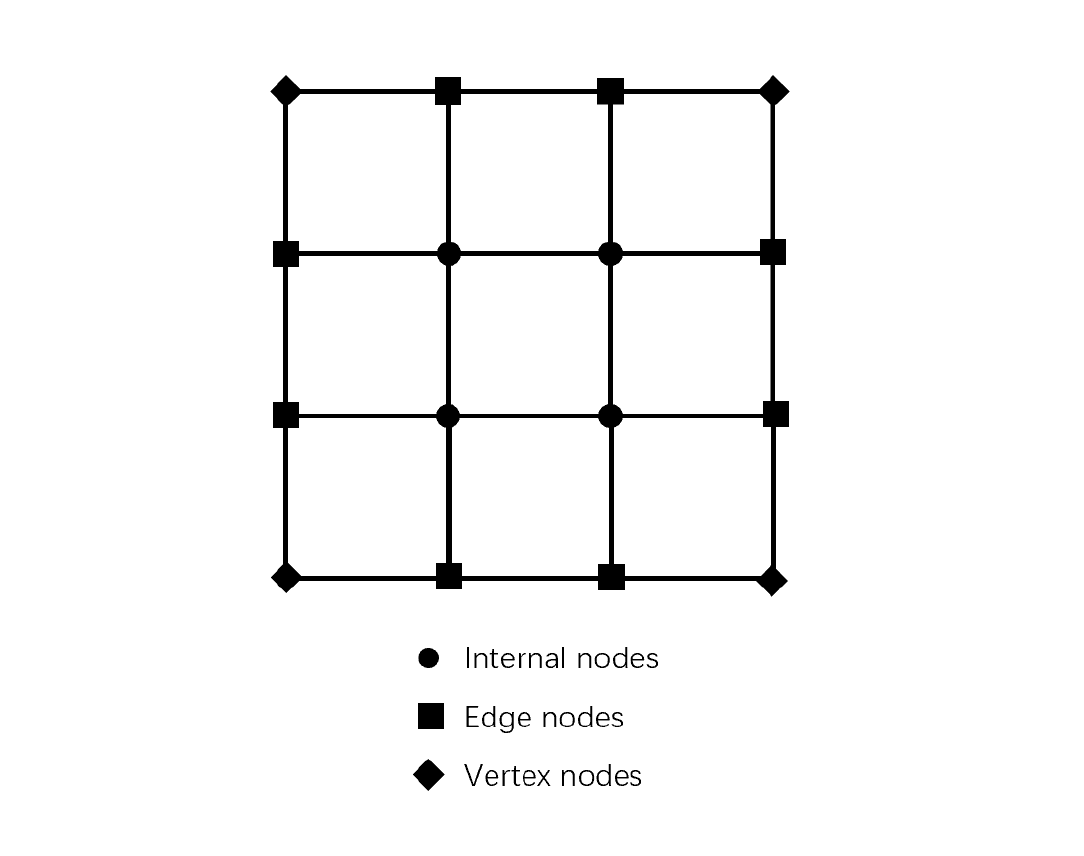}
	\captionof{figure} {Illustration of the 3 categories of Internal, Edge, and Vertex cells, corresponding to the position of each fine cell within a coarse element.}\label{localdomain_ill}
\end{center}

Note that the vertex nodes are in fact the coarse mesh nodes, where the coarse-scale solution will be computed. The basis functions are needed to interpolate the solution between the vertex cells through the edge and internal cells. 

To develop the basis functions, first the fine-scale stiffness matrix $K^h$ is permuted, such that the terms for vertex, then edge and finally the internal cells appear. The permutation operator $\mathbf{T}$ as such reorders $K^h$ into $K^v$such that
\begin{gather}
\Breve{K}^v = \mathbf{T}  K^h \mathbf{T}^T = 
\mathbf{T} \ 
\begin{bmatrix}
   {K}_{uu} & {K}_{ua} & {K}_{ub} \\  
   {K}_{au} & {K}_{aa} & {K}_{ab} \\ 
   {K}_{bu} & {K}_{ba} & {K}_{bb} 
   \end{bmatrix} \ 
   \mathbf{T}^T
=
\begin{bmatrix} 
   {K}_{II} & {K}_{IE} & {K}_{IV} \\  
   {K}_{EI} & {K}_{EE} & {K}_{EV} \\ 
   {K}_{VI} & {K}_{VE} & {K}_{VV} 
   \end{bmatrix}.
\end{gather}
Here, $I$ represents the internal nodes, $E$ represents the edge nodes and $V$ represents the vertex nodes. The permuted linear system, therefore, reads
\begin{gather}
\begin{bmatrix} 
  {K}_{II} & {K}_{IE} & {K}_{IV} \\
  {K}_{EI} & {K}_{EE} & {K}_{EV} \\
  {K}_{VI} & {K}_{VE} & {K}_{VV}
  \end{bmatrix}
\begin{bmatrix}
{d}_I \\
{d}_E \\
{d}_V
\end{bmatrix}
=
\begin{bmatrix}
{f}_I \\
{f}_E \\
{f}_V
\end{bmatrix}
\end{gather}

Note that the permuted system collects all entries of the XFEM discrete system belonging to I, E, and V cells. Therefore, the XFEM enrichment entries due to tips and jumps are within their corresponding I, E, and V entries.

The reduced-dimensional boundary condition is now being imposed by replacing the 2D equation for E by a 1D XFEM discrete system. This leads the entry $\bar{K}_{VI}$ to vanish, as there will be no connectivity between the edge and internal cells for the edge cells. This 1D edge equations can then be expressed as
\begin{equation}
{K}_{EE}^R {d}_E+{K}_{EV}^R {d}_V = 0.
\end{equation}

Knowing that the solution at vertex cells will be obtained from the coarse-scale system, the reorder fine-scale system matrix can now be reduced to
\begin{gather}\label{reduced_bs}
\begin{bmatrix} 
  {K}_{II} & {K}_{IE} & {K}_{IV} \\
  0 & {K}_{EE}^R & {K}_{EV}^R \\
  0 & 0 & I_{VV}
  \end{bmatrix}
\begin{bmatrix}
{d'}_I \\
{d'}_E \\
{d'}_V
\end{bmatrix}
=
\begin{bmatrix}
0 \\
0 \\
0
\end{bmatrix}.
\end{gather}
Note that the equations for basis functions do not have any source terms in their right-hand-side. The upper-triangular matrix of Eq. \eqref{reduced_bs} can be easily inverted to give the prolongation operator, i.e., given the coarse nodes solutions $d'_V$, one can obtain the solution at the edge via
\begin{equation}
{d'}_E = - (K^R_{EE})^{-1} {K^R}_{EV} \ {d}_V' = P_E \ {d}_V'.
\end{equation}
similarly, the solution at the internal cells reads
\begin{align}
{d'}_I &= - K_{II}^{-1} (K_{IE} d'_E + K_{IV} d'_V) \nonumber\\
       &=  - K_{II}^{-1} (- K_{IE} (K^R_{EE})^{-1} {K^R}_{EV} + K_{IV}) \ d'_V = P_I \ {d}_V'.
\end{align}
Note that $P_E$ and $P_I$ are the sub-matrices of the prolongation operator, i.e.,
\begin{equation}
   d' = 
   \begin{bmatrix}
     d'_I\\
     d'_E\\
     d'_V
   \end{bmatrix}
   =
   \underbrace{
   \begin{bmatrix}
     - K_{II}^{-1} (- K_{IE} (K^R_{EE})^{-1} {K^R}_{EV} + K_{IV})\\
     - (K^R_{EE})^{-1} {K^R}_{EV}\\
     I_{VV}
   \end{bmatrix}}_{\mathbf{P}}
   \ d'_V.
\end{equation}
Here, $I_{VV}$ is the diagonal identity matrix equal to the size of the vertex nodes. 

After defining the prolongation operator algebraically, based on the entries of the 2D XFEM (for internal cells) and 1D XFEM (for edge cells), one can find the multiscale solution.

\subsection{Iterative multiscale procedure (iMS-XFEM)}
The multiscale solutions with the accurate XFEM basis functions can be used to provide an approximate efficient solution for many practical applications. However, it is important to control the error and reduce it to any desired tolerance \cite{Multiscale2_Hadi} if needed. As such, the MS-XFEM is paired with a fine-scale smoother (here, ILU(0) \cite{Chow1997}) to allow for error reduction. Note that this iterative procedure can also be used within a GMRES iterative loop \cite{Saad86} to enhance convergence rates. The study of the most efficient iterative strategy to reduce the error is outside the scope of the current paper. The iterative procedure then reads

\begin{itemize}
    \item Construct the $\mathbf{P}$ and $\mathbf{R}$ operators
    \item Iterate until $||r^{\nu+1}|| = ||f^h - K^h d'^{\nu+1}|| < e_r$
    \begin{itemize}
        \item [1.] MS-XFEM stage: $\delta{d'}^{\nu+1/2} = \mathbf{P} (\mathbf{R} K^h \mathbf{P})^{-1} \mathbf{R} \ r^{\nu} $
        \item [2.] Smoothing stage (apply $n_s$ times ILU(0)): $\delta{d'}^{\nu+1} = ({M^{ns}_{\text{ILU(0)}}})^{-1} \ r^{\nu+1/2}$ 
    \end{itemize}
\end{itemize}
Note that $n_s$ is defined by user. 

\section{Numerical Test Cases}   
In this section several test cases are considered to investigate the performance of MS-XFEM both as approximate solver and integrated within the iterative error reduction procedure. 

\subsection{Test case 1: Single fracture in a heterogeneous domain}
\begin{figure}[H]
	\centering
	\begin{subfigure}{0.4\textwidth}
		\centering
		\includegraphics[trim={6cm 18cm 6cm 4cm}, clip, width=\textwidth]{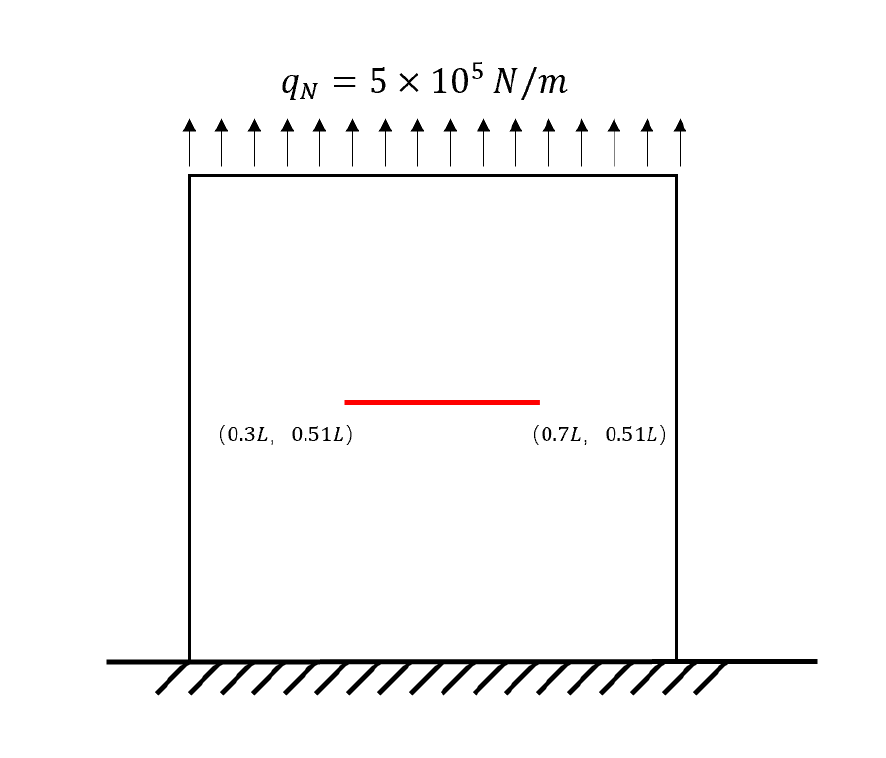}
		\caption{}
	\end{subfigure}
	\begin{subfigure}{0.4\textwidth}
	\centering
	\includegraphics[trim={4cm 8cm 4cm 8cm}, clip, width=\textwidth]{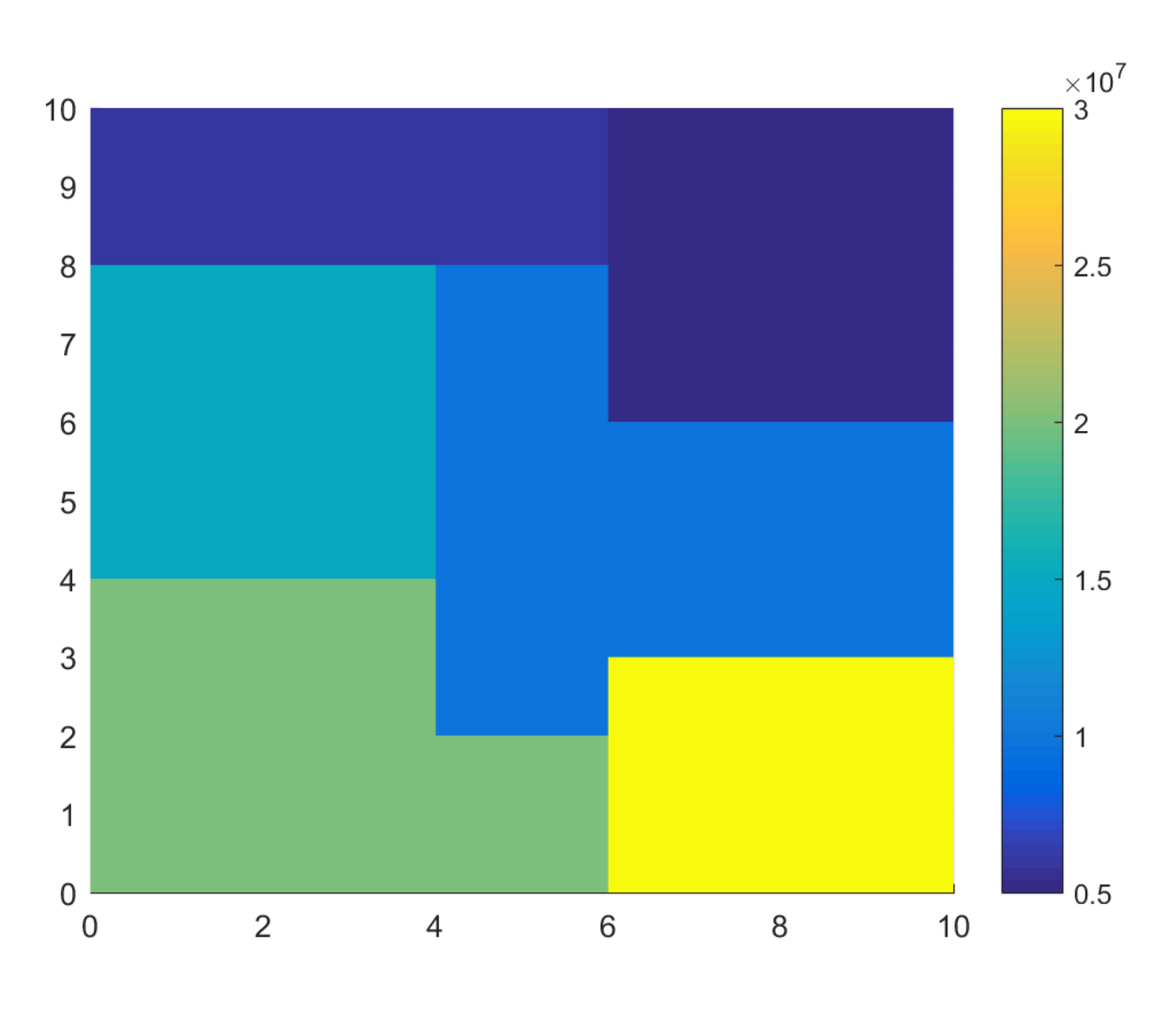}
	\caption{}
	\end{subfigure}
	\caption {Test case 1 :(a) illustration of the model setup, (b) heterogeneous Young's modulus distribution. Note the units are SI.}\label{Test1_f1}
\end{figure}

In the first test, a square 2D domain of $L \times L$ with $L=10$[m] is considered, which contains a single horizontal fracture in its centre, as shown in figure \ref{Test1_f1}a. The fine-scale mesh consists of $40\times 40$ cells, while the MS-XFEM contains only $5\times 5$ coarse grids. This results in a coarsening ratio of 8, in each direction. The heterogeneous Young's modulus distribution is shown in figure \ref{Test1_f1}b, while the Possion's ratio is assumed to be constant 0.2 everywhere. The fracture tip coordinates are shown in figure \ref{Test1_f1}a. The Dirichlet boundary condition is set at the south face, while the north boundary is under distributed upward load with $q = 5 \times 10^{5}$ [N/m] magnitude.

\begin{figure}[H]
	\centering
	\begin{subfigure}{0.49\textwidth}
		\centering
		\includegraphics[trim={5cm 9cm 4cm 8cm}, clip, width=0.8\textwidth]{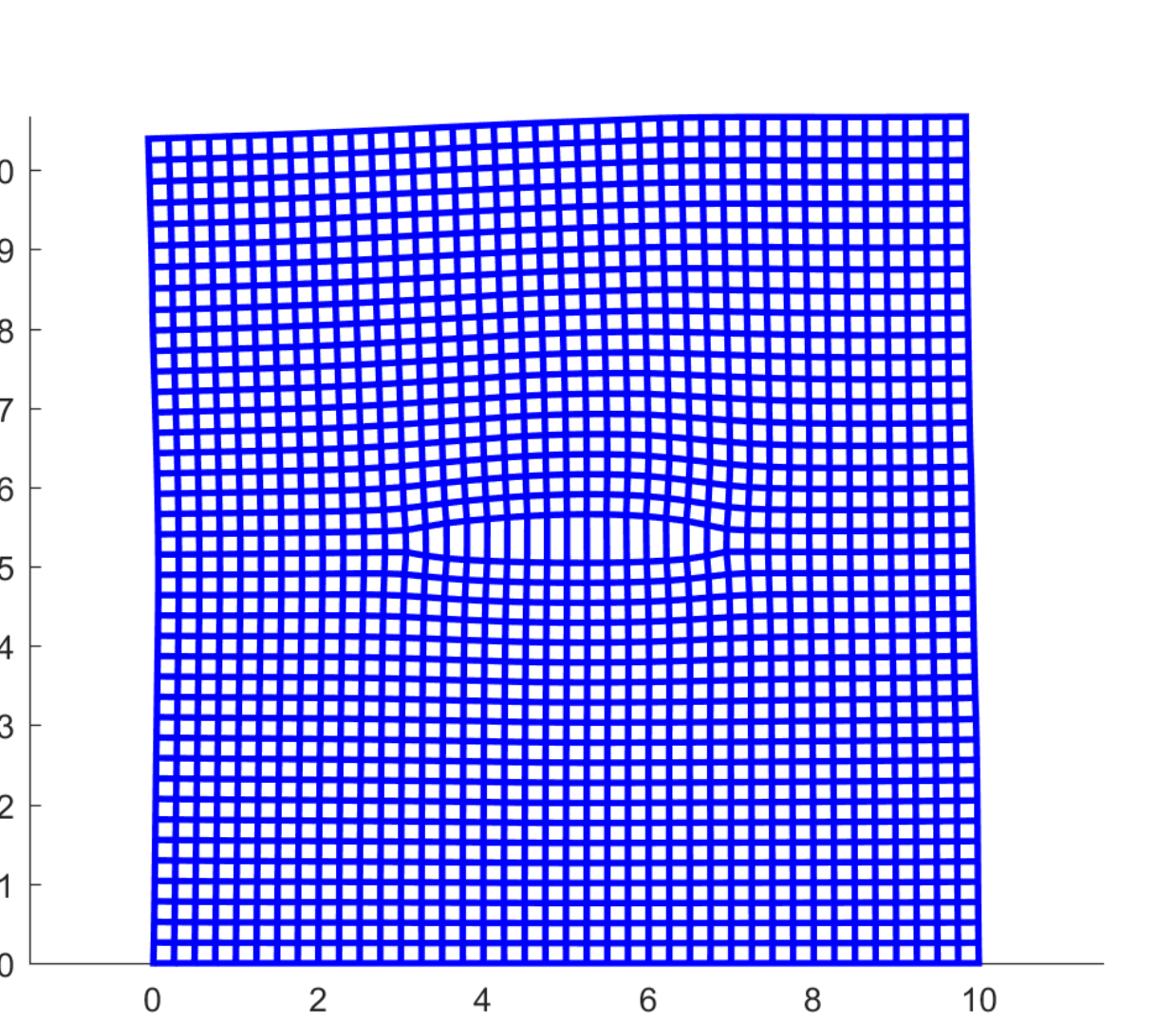}
		\caption{}
	\end{subfigure}\\
	\begin{subfigure}{0.49\textwidth}
		\centering
		\includegraphics[trim={5cm 9cm 4cm 8cm}, clip, width=0.8\textwidth]{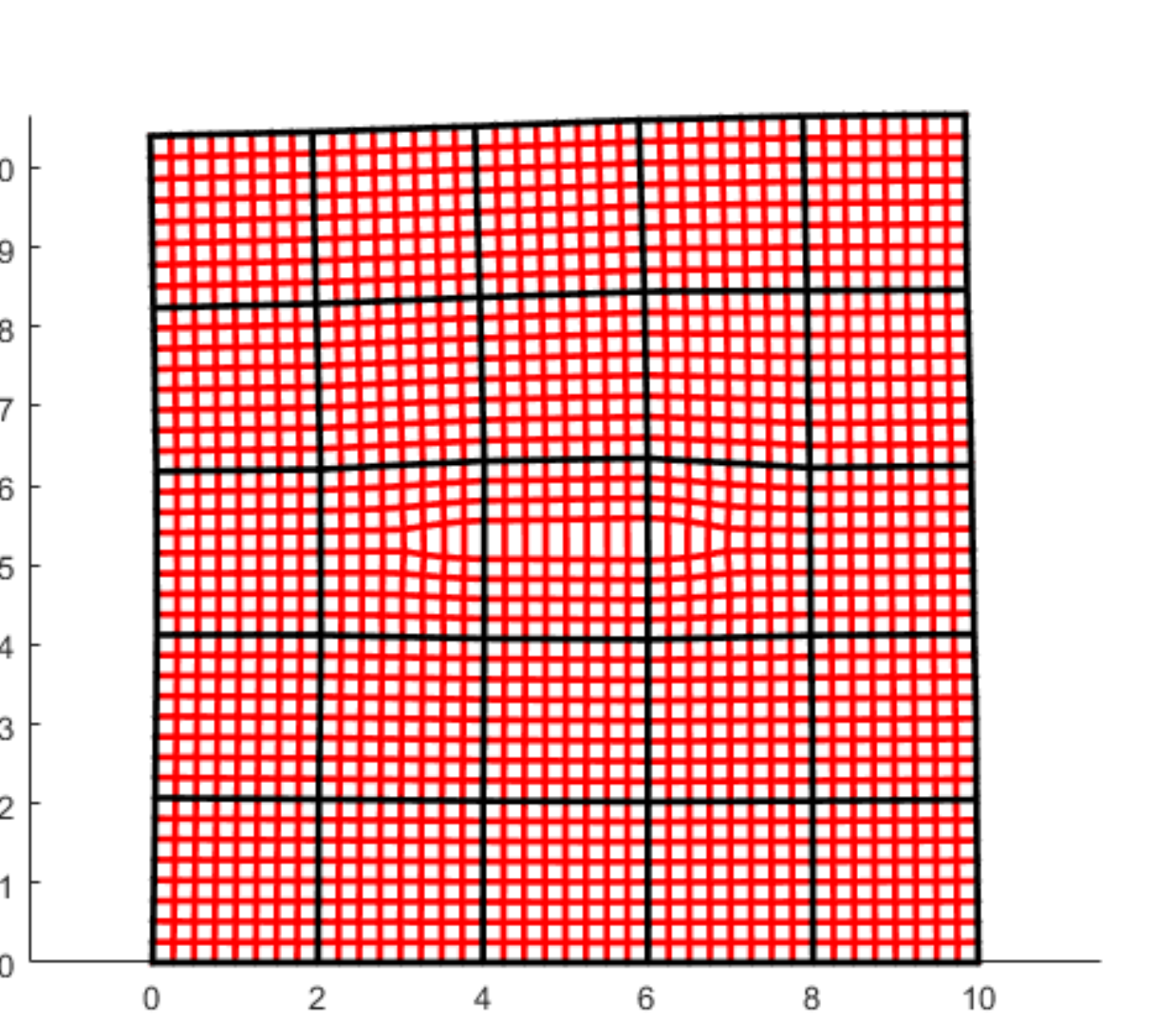}
		\caption{$||e_y||=2.6512 \times 10^{-4}$}
	\end{subfigure}
	\begin{subfigure}{0.49\textwidth}
		\centering
		\includegraphics[trim={5cm 9cm 4cm 8cm}, clip, width=0.8\textwidth]{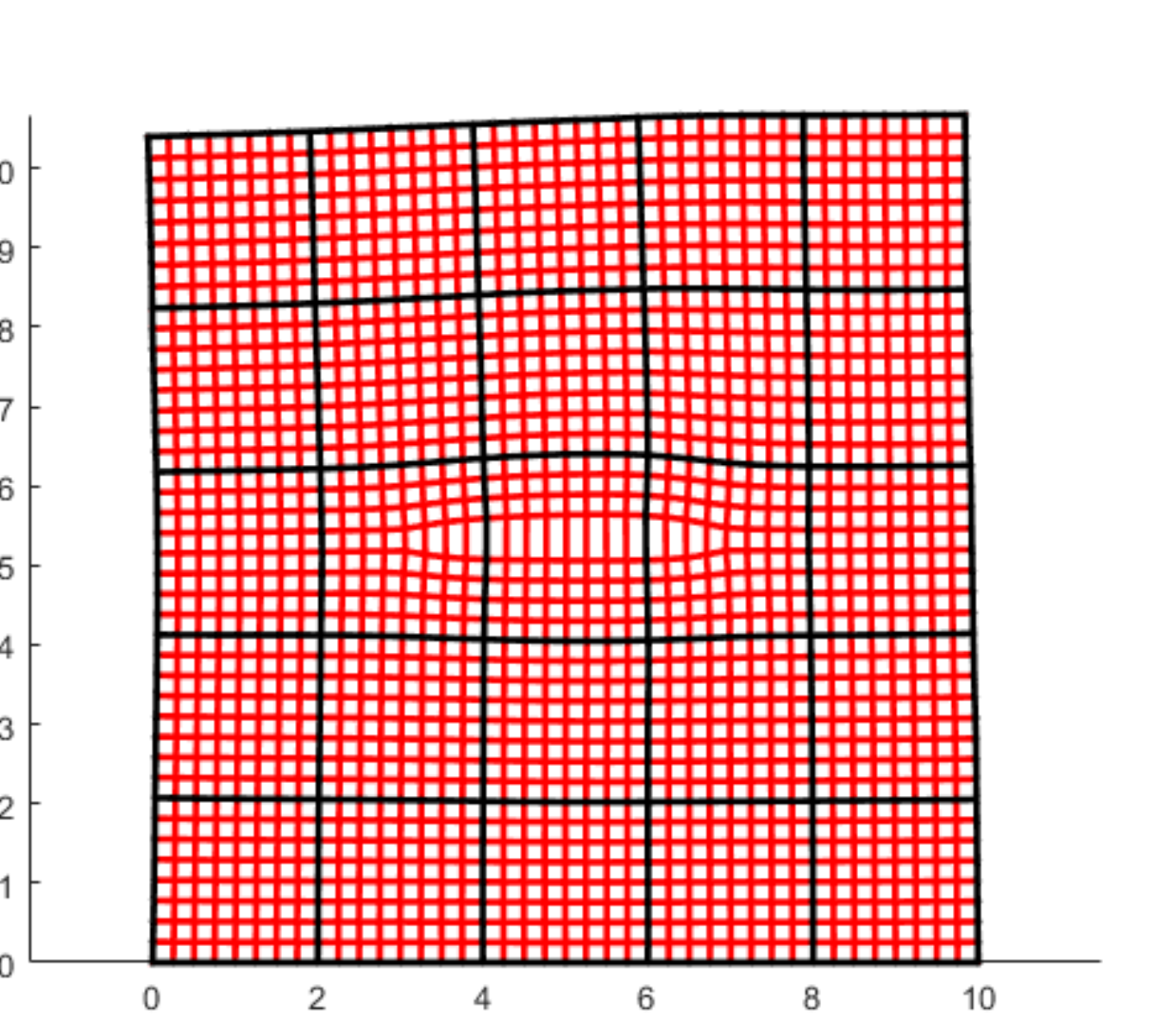}
		\caption{$||e_y||=1.0172 \times 10^{-4}$}
	\end{subfigure}
	\caption {Test Case 1: displacement field for a heterogeneous fractured reservoir using (a) fine scale XFEM and (b) MS-XFEM (c) iMS-XFEM after 3 iterations.}\label{Test1_f1_r}
\end{figure}

Results are shown in figure \ref{Test1_f1_r}. The black lines on figure \ref{Test1_f1_r} (b) and (c) represent the coarse scale mesh. It is clear that the results of MS-XFEM on only $5 \time 5$ grid cells is in reasonable agreement with that of the fine-scale XFEM solver using a $40 \times 40$ mesh. Note that no enrichment for the MS-XFEM is used, and the basis functions are computed using the XFEM method on local domains. 

\begin{figure}[H]
	\centering
	\begin{subfigure}{0.49\textwidth}
		\centering
		\includegraphics[trim={4cm 17cm 4cm 3cm}, clip, width=0.8\textwidth]{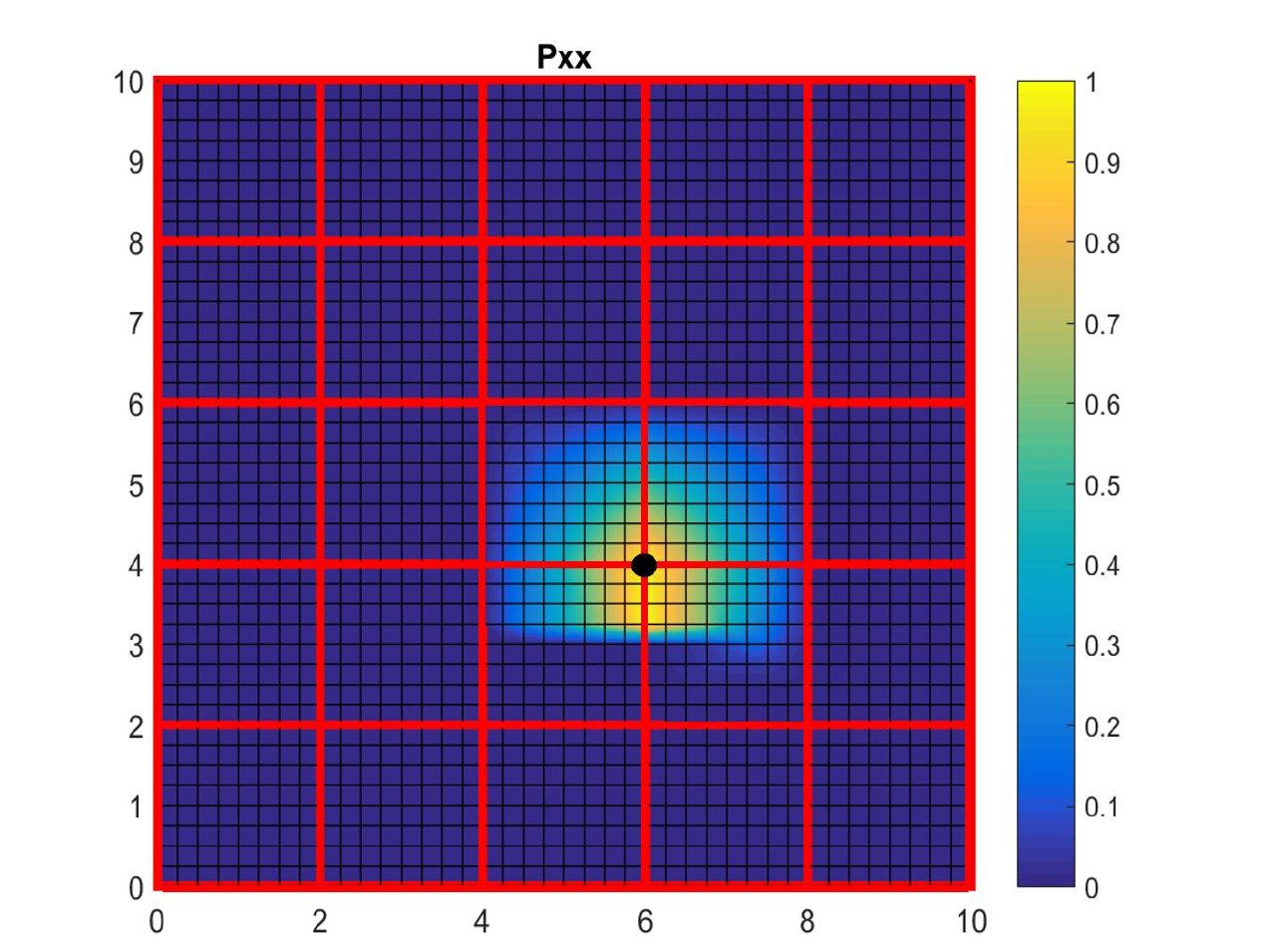}
		\caption{}
	\end{subfigure}%
	\begin{subfigure}{0.49\textwidth}
		\centering
		\includegraphics[trim={4cm 17cm 4cm 3cm}, clip, width=0.8\textwidth]{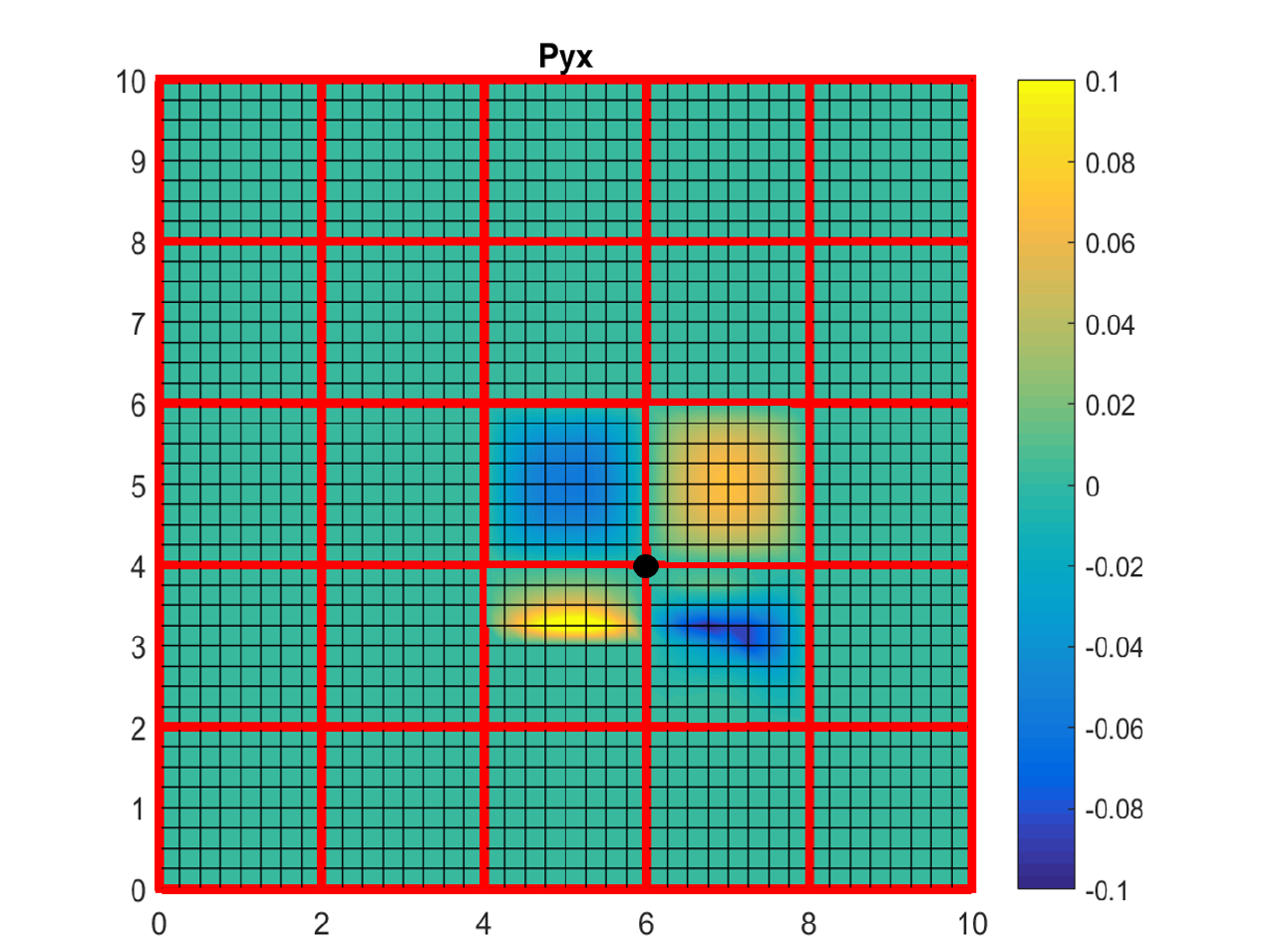}
		\caption{}
	\end{subfigure}
	\caption {Basis functions of single fracture test case. Single discontinuity is captured by axial equilibrium and transverse equilibrium solutions}\label{Test1_basis}
\end{figure}

A basis function for a fractured local domain is illustrated in figure \ref{Test1_basis}. Note that the discontinuity is captured by the basis functions, since XFEM is used to solve for it. 

The effect of the coarsening ratio is shown in figure \ref{Test1_cr}. The error $e$ in this figure is computed using 
\begin{equation}
e_i=\frac{||u_{i,MS}-u_{i,f}||_2}{N}, \ \ \ \forall i\in{x,y},
\end{equation}
where N is the number of fine-scale mesh nodes. $u_{i,MS}$ and $u_{i,f}$ denote the proloned MS-XFEM solution displacement field and fine-scale solution displacement field, respectively.

\begin{center}
	\centering
	\includegraphics[trim={4cm 8cm 4cm 8cm}, clip, width=0.5\textwidth]{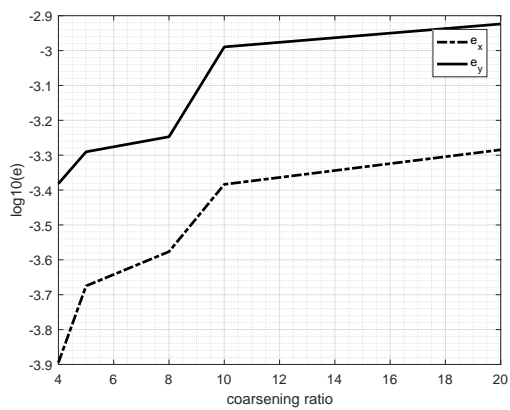}
	\captionof{figure} {Change of errors with different coarsening ratios}\label{Test1_cr}
\end{center}

The MS-XFEM errors are due to the local boundary conditions used to calculate the basis functions, and also because no additional enrichment functions are imposed at coarse scale. Note this means that for heterogeneous domains there can be a finer resolution for coarse cells at which the local boundary conditions impose more errors compared with coarser resolutions. In spite of this, figure \ref{Test1_cr} clearly shows, for this example, a decaying trend of the error with respect to the finer coarse mesh is observed.

\subsubsection{Iterative MS-XFEM}
\begin{figure}[H]
	\centering
	\begin{subfigure}{0.49\textwidth}
		\centering
    	\includegraphics[trim={4cm 8cm 4cm 9cm}, clip, width=\textwidth]{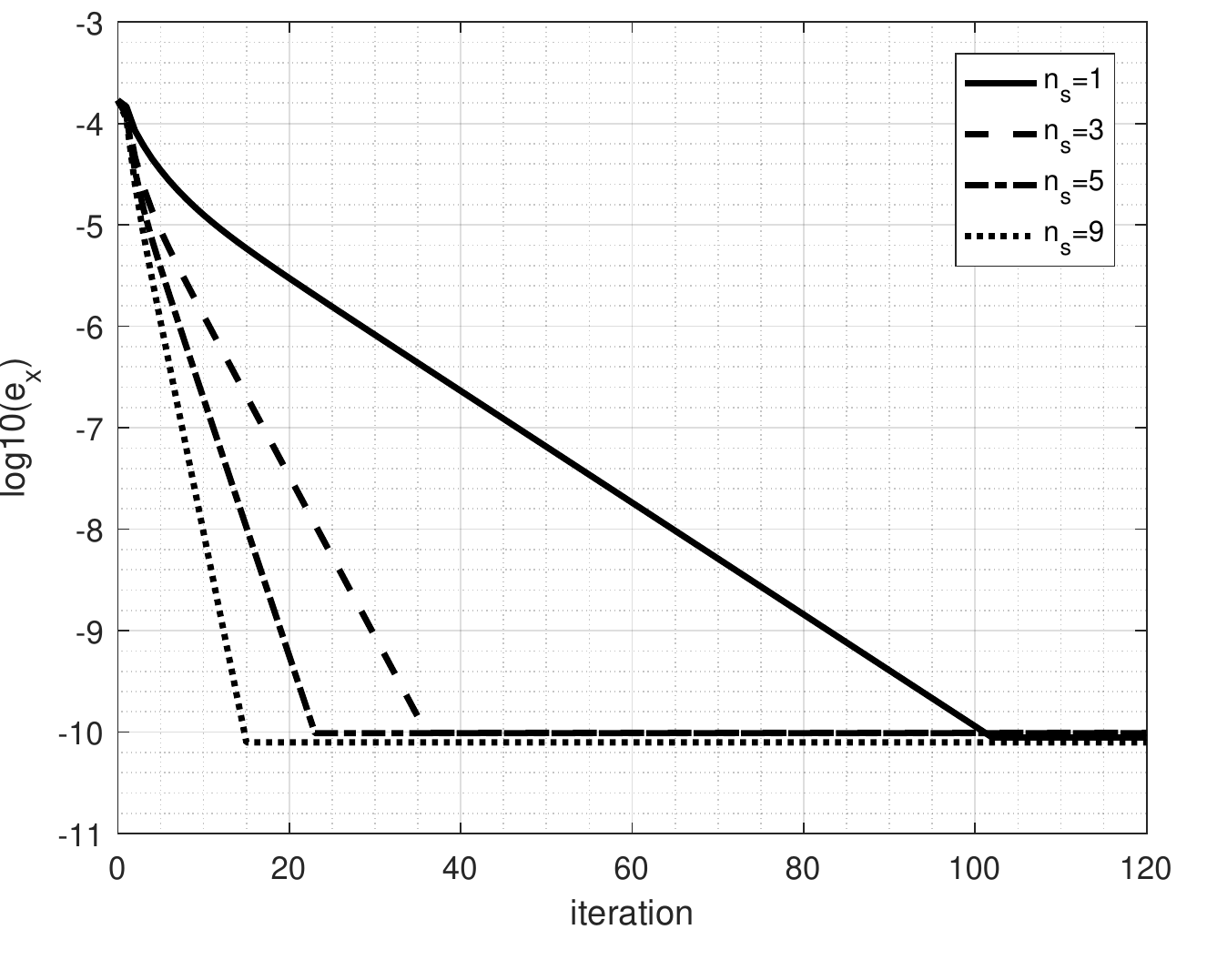}
		\caption{}
	\end{subfigure}
	\begin{subfigure}{0.49\textwidth}
	\centering
	\includegraphics[trim={4cm 8cm 4cm 9cm}, clip, width=\textwidth]{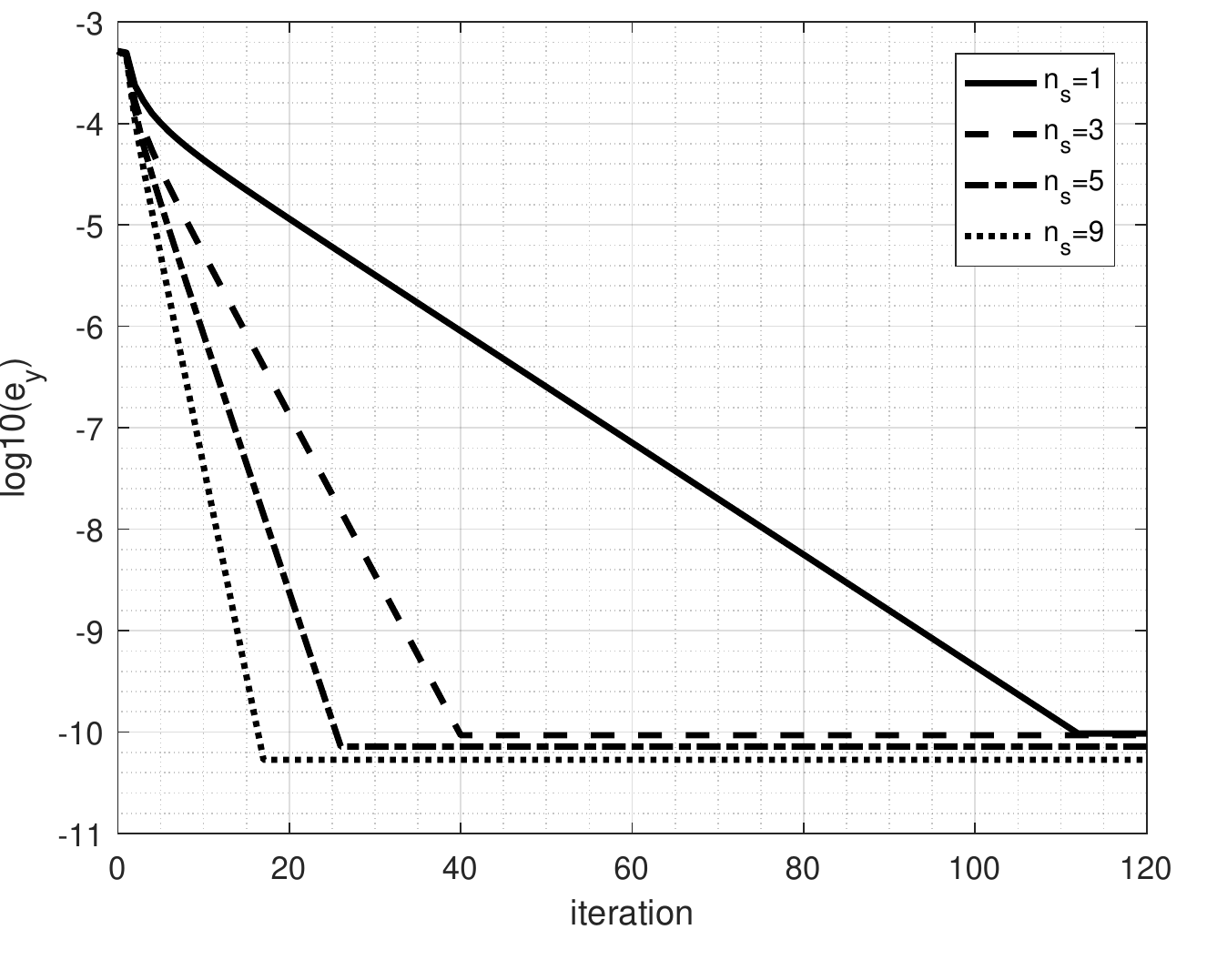}
	\caption{}
	\end{subfigure}
	\caption {Iteration history of iMS-XFEM procedure with different number of smoothings per step. Errors for displacement in x (a) and y (b) direction are shown.}\label{Test1_imsxfem}
\end{figure}

As discussed in section 3.4, one can pair the MS-XFEM in an iterative strategy in which the error is reduced to any desired level \cite{YWang2014}. From figure \ref{Test1_f1_r} (c) that with 3 times of fine scale smoothers applied in the second stage after 3 iterations the MS-XFEM result has been improved compared to the fine-scale result and the error is decreased significantly.  Results of the iterative MS-XFEM procedure (iMS-XFEM) are shown in figure \ref{Test1_imsxfem}. Different smoothing steps per iteration values $n_s$ are used. Note than neither GMRES \cite{Saad86} nor any other iterative convergence enhancing procedure is used here. Clearly, one can reduce the multiscale errors to the machine accuracy by applying iMS-XFEM iterations. In particular, for practical applications, one can stop iterations after a few counts, once the error norm is below the level of uncertainty $\tau$ within the parameters of the problem.

\begin{equation}
e_{i}\leqslant\tau, \ \ \ i={x,y}
\end{equation}
In which $\tau$ is chosen as $10^{-10}$ in here.\\
In figure 12 it is shown that convergence is achieved with $n_s$ rounds of the fine scale smoother involved in the second stage.\\

\subsection{Test case 2: heterogeneous reservoir with multiple fractures}
 
 The second test case is set to  model deformation in a heterogeneous reservoir with more fractures. The size and the heterogeneous properties of this test case are the same as those in test case 1. Here, more fractures are considered. In addition, compared to test case 1, the east and west boundaries are also observing distributed loads, as shown in figure \ref{test2_1}.
\begin{center}
	\centering
	\includegraphics[trim={6cm 18cm 6cm 4cm}, clip, width=0.5\textwidth]{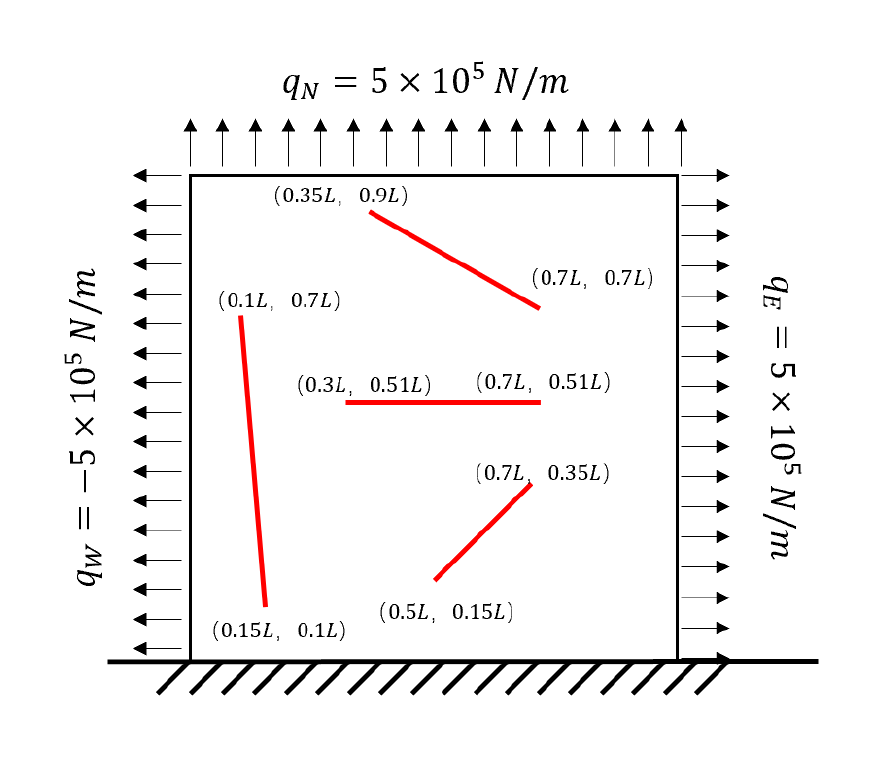}
	\captionof{figure} {Test case 2: Multiple fractures within a heterogeneous reservoir under tension stress across three boundaries.}\label{test2_1}
\end{center}

\begin{figure}[H]
	\centering 
	\begin{subfigure}{0.33\textwidth}
		\centering
		\includegraphics[trim={5cm 9cm 4cm 8cm}, clip, width=0.9\textwidth]{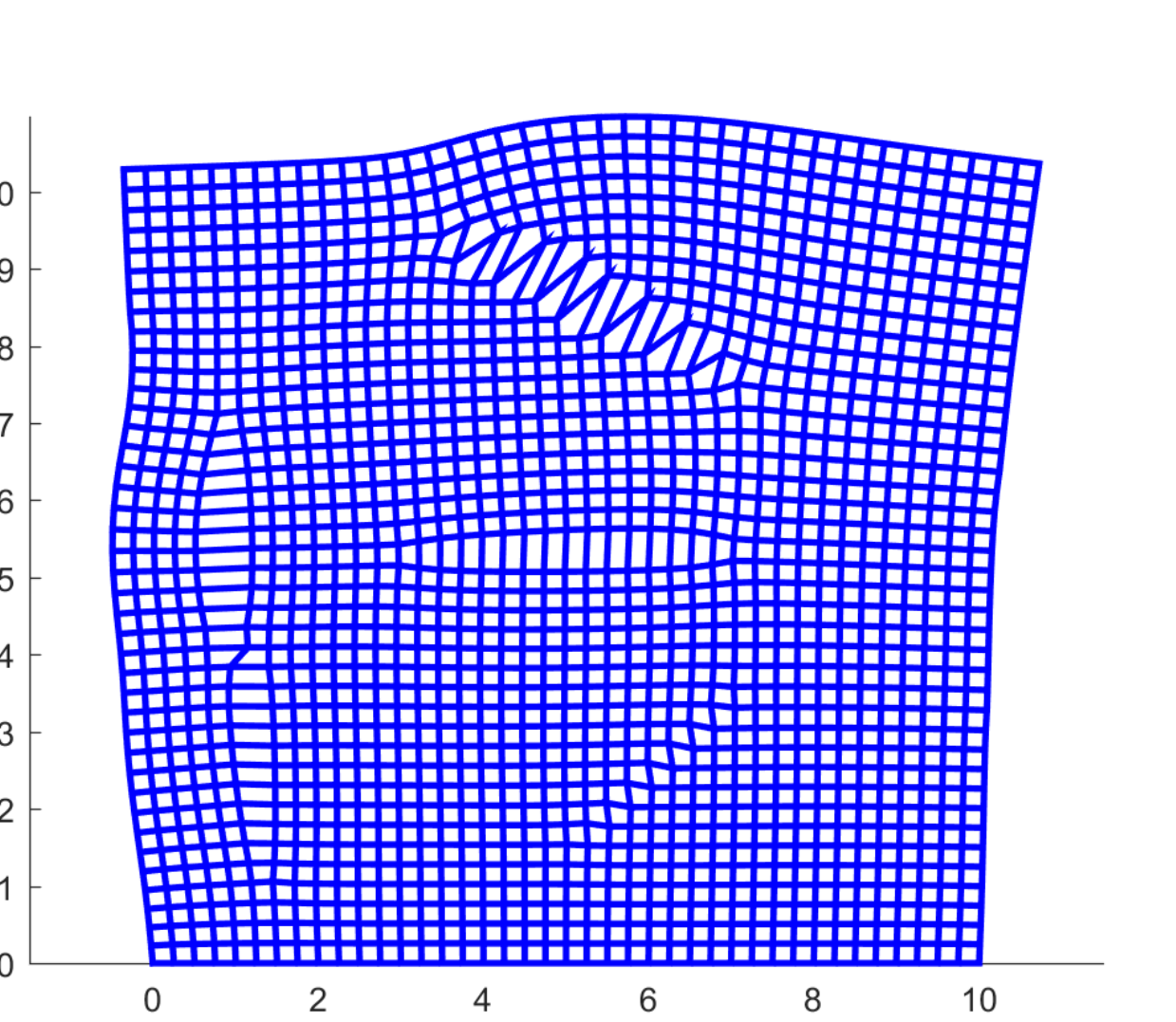}
		\caption{}
	\end{subfigure}\\
	\begin{subfigure}{0.33\textwidth}
		\centering
		\includegraphics[trim={5cm 9cm 4cm 8cm}, clip, width=0.9\textwidth]{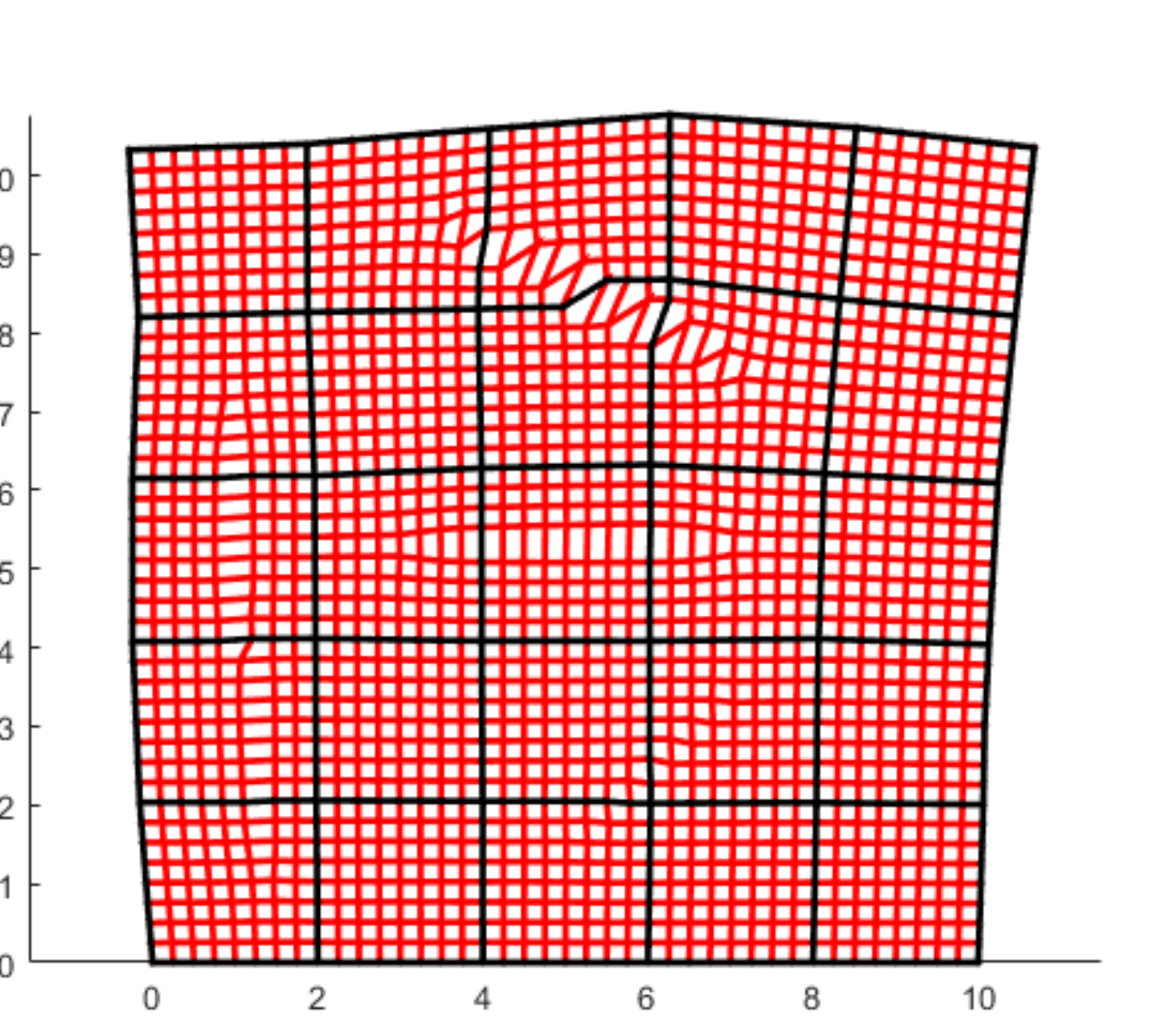}
		\caption{$||e_y||=1.5 \times 10^{-3}$}
	\end{subfigure}
		\begin{subfigure}{0.33\textwidth}
		\centering
		\includegraphics[trim={5cm 9cm 4cm 8cm}, clip, width=0.9\textwidth]{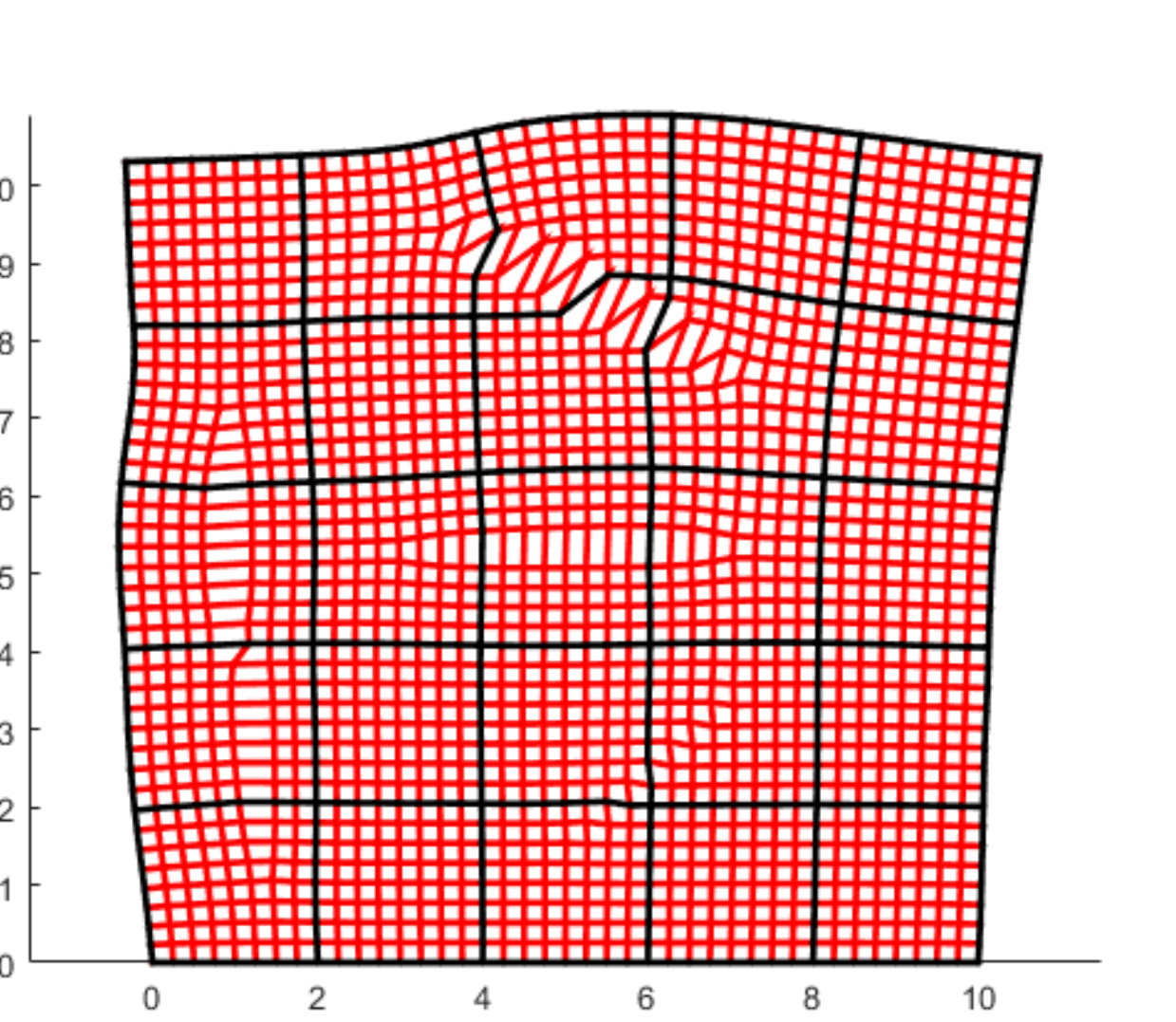}
		\caption{$||e_y||=4.0566 \times 10^{-4}$}
	\end{subfigure}
	\caption {Test Case 2: displacement field for a heterogeneous media with multiple fractures for (a) fine scale XFEM and (b) MS-XFEM without iterative strategy applied (c) iMS-XFEM result after 3 iterations.}\label{test2_2}
\end{figure}

Simulation results for both fine-scale XFEM and MS-XFEM are shown in figure \ref{test2_2}. The black lines on figure \ref{test2_2} (b) and (c) represent the coarse scale mesh. It is clear that the MS-XFEM (without using iterations) results in a relatively accurate representation of the deformation field, compared with the fine-scale XFEM, using $8 \times 8$ fewer grid cells and no coarse-scale enrichment functions. 

\begin{figure}[H]
	\centering
	\begin{subfigure}{0.49\textwidth}
		\centering
		\includegraphics[trim={4cm 17cm 4cm 3cm}, clip, width=0.8\textwidth]{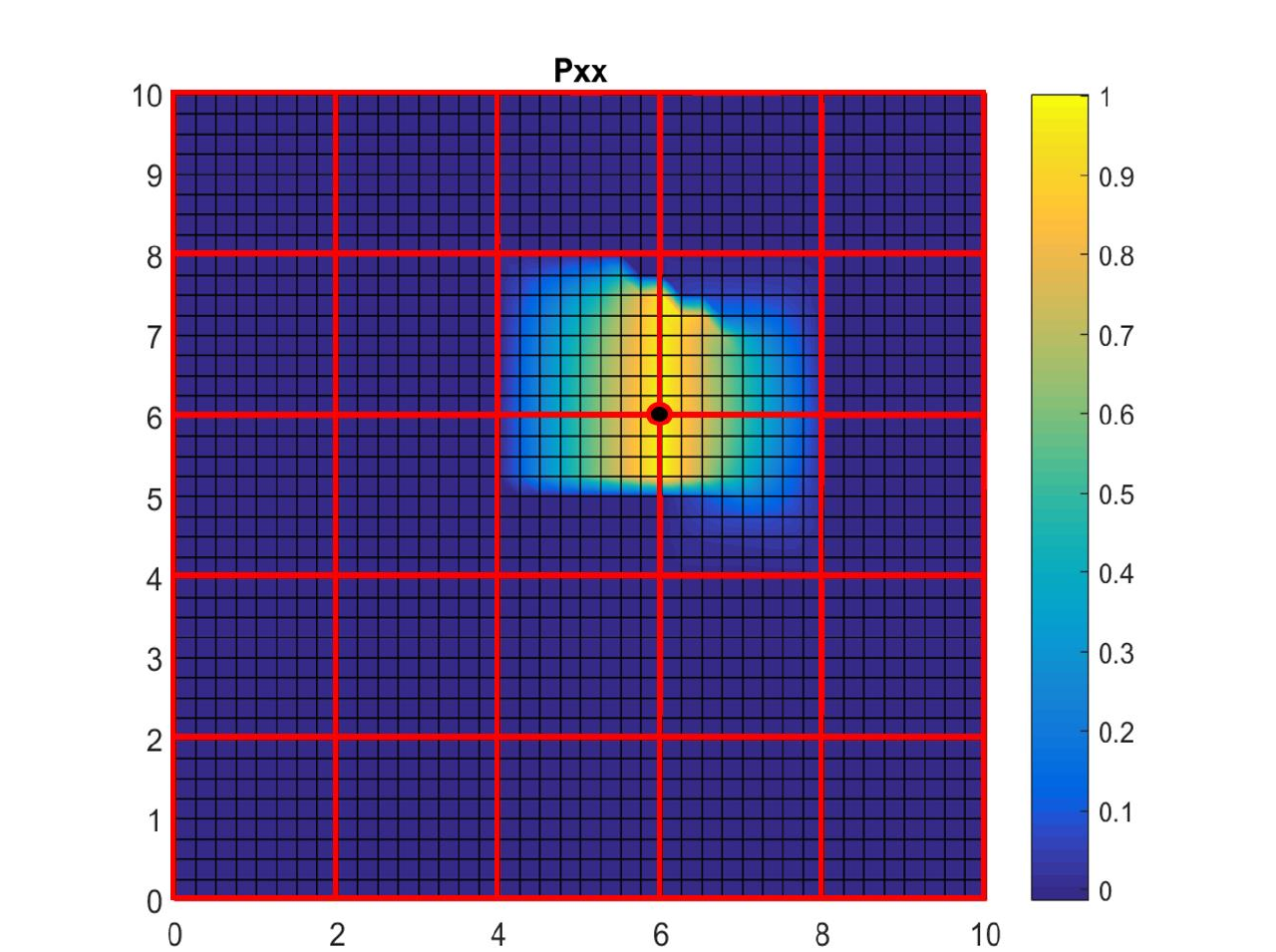}
		\caption{}
	\end{subfigure}%
	\begin{subfigure}{0.49\textwidth}
		\centering
		\includegraphics[trim={4cm 17cm 4cm 3cm}, clip, width=0.8\textwidth]{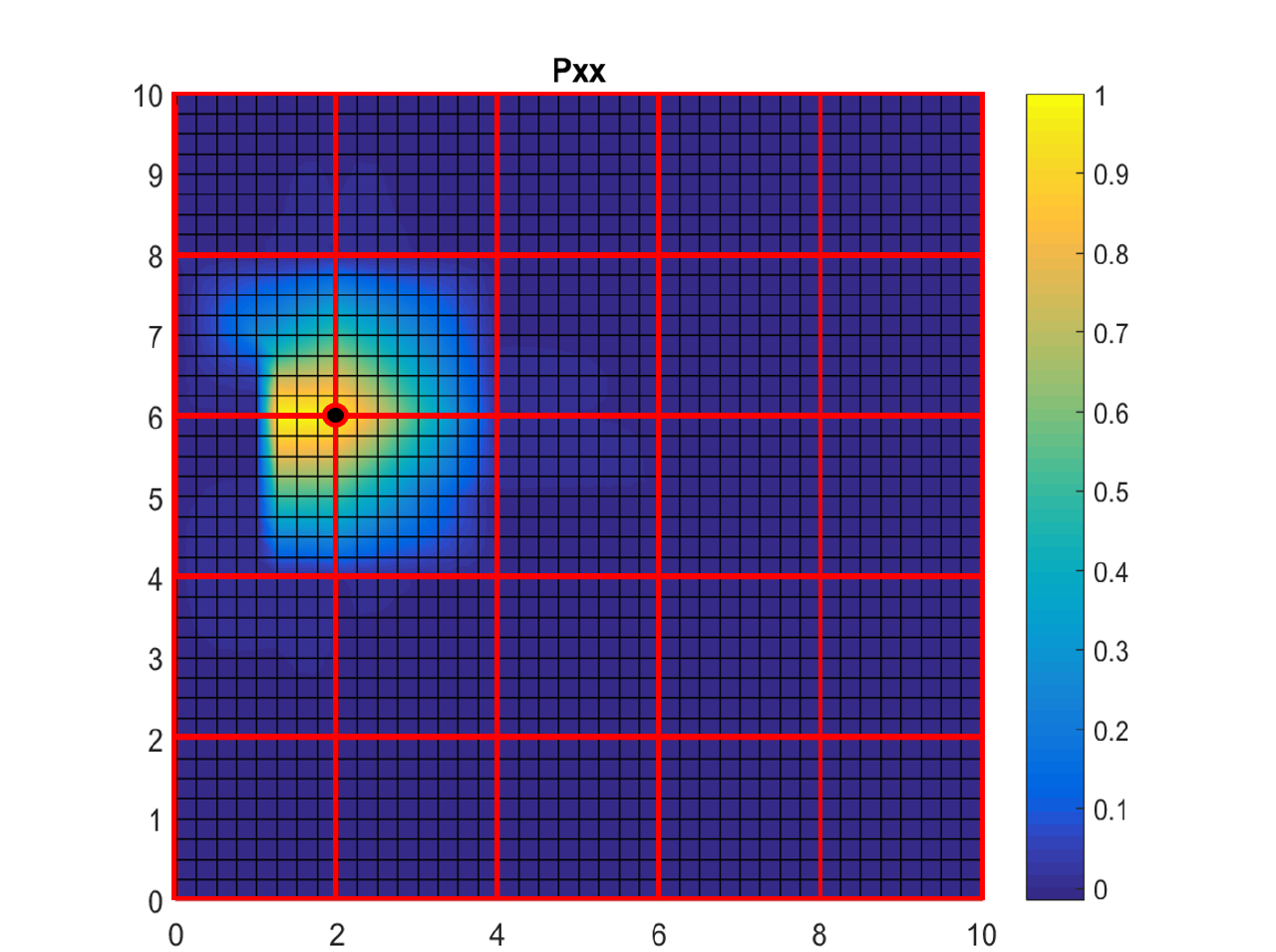}
		\caption{}
	\end{subfigure}
	\caption {Illustration of the basis functions $P_{xx}$ for two different part of the domain, where two (a) and one (b) discontinuities are captured.}\label{test2_3}
\end{figure}

An example of two basis functions for this test case is shown in figure \ref{test2_3}. In the local plot, it is illustrated how 1 (\ref{test2_3}a) and 2 (\ref{test2_3}b) fractures are captured by the basis functions. 

\begin{figure}[H]
	\centering
	\begin{subfigure}{0.49\textwidth}
		\centering
		\includegraphics[trim={4cm 8cm 4cm 9cm}, clip, width=\textwidth]{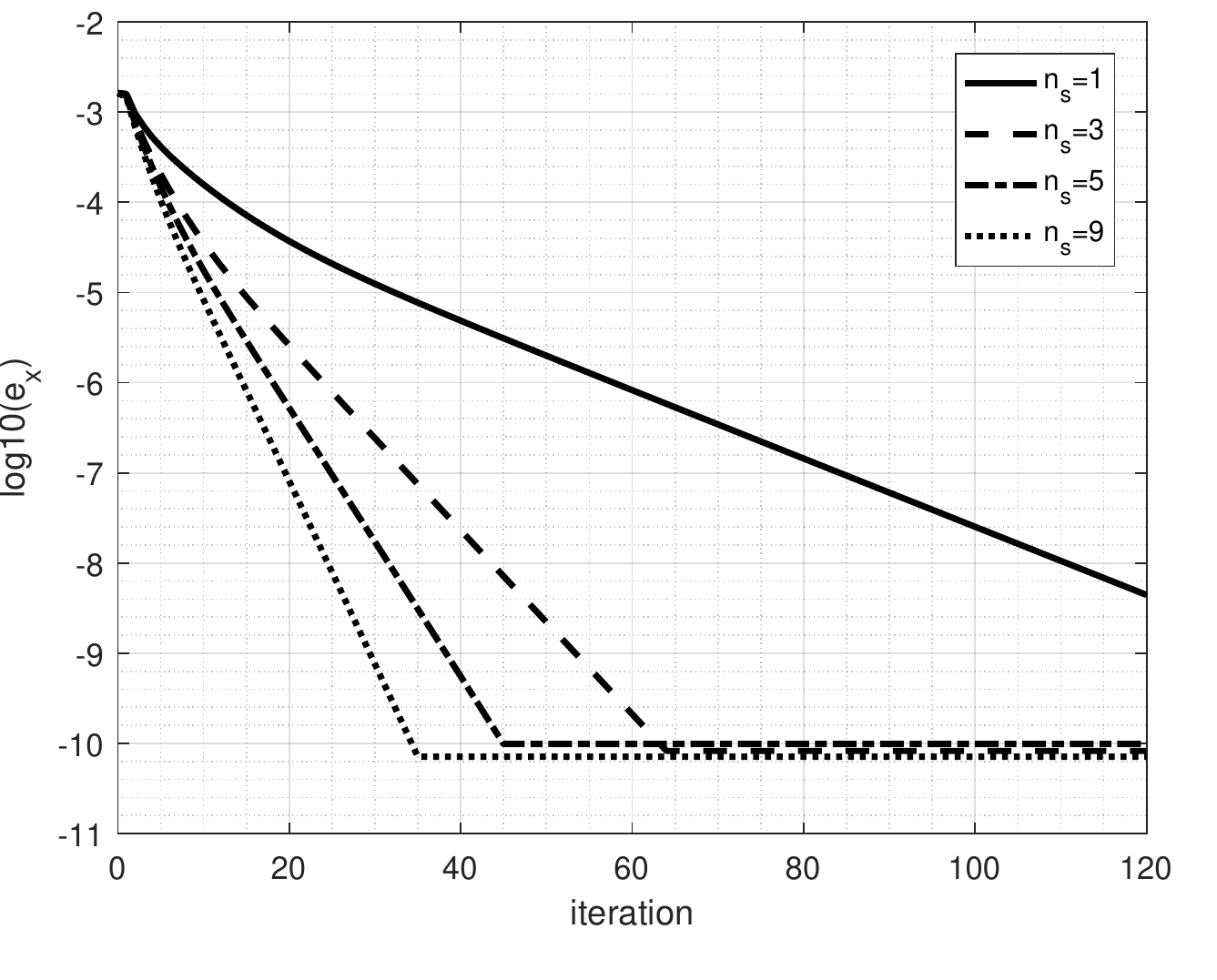}
		\caption{}
	\end{subfigure}%
	\begin{subfigure}{0.49\textwidth}
		\centering
		\includegraphics[trim={4cm 8cm 4cm 9cm}, clip, width=\textwidth]{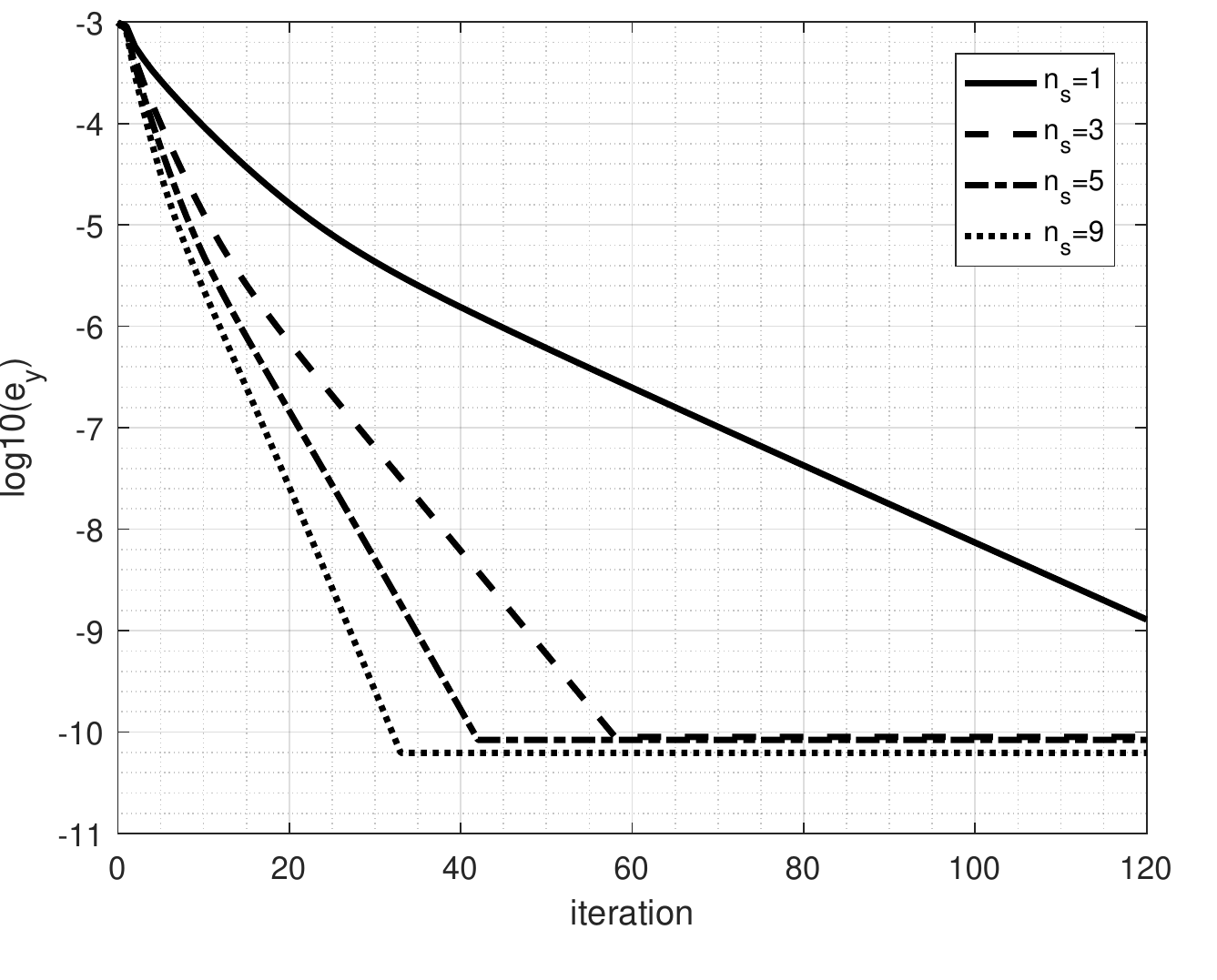}
		\caption{}
	\end{subfigure}
	\caption {Iteration history of iMS-XFEM procedure with different number of smoothing per step. Errors for displacement in x (a) and y (b) direction are shown.}\label{test2_4}
\end{figure}

The iMS-XFEM procedure, as explained before, is now applied to reduce the multiscale errors to machine precision. Still in figure \ref{test2_2} (c) the result quality is improved a lot after 3 iterations with 3 times fine-scale smoothers applied in the second stage. Note that since no GMRES nor a complete smoother is used, but the incomplete smoother ILU(0) for its efficiency. Results are shown in figure \ref{test2_4}.

\section{Conclusion}\label{sec:conclusion}
A multiscale procedure for XFEM is proposed to model deformation of geological heterogeneous fractured fields. The method resolves the discontinuities through local multiscale basis functions, which are computed using XFEM subjected to local boundary conditions. 
The coarse-scale system is obtained by using the basis functions, algebraically, which does not have any additional enrichment functions, in contrast to the local basis function systems. This procedure makes the MS-XFEM very efficient. Also, by combining it with a fine-scale smoother, an iterative MS-XFEM (iMS-XFEM) procedure is developed, which allows to reduce the error to any desired level of accuracy.

Two heterogeneous test cases were studied as proof-of-concept, to investigate the performance of the MS-XFEM. It was shown that MS-XFEM results in acceptable solutions, when no iterations are imposed. By applying iterations, one can further improve the results. For practical applications, when parameters are uncertain, only a few iterations can be applied to maintain (and control) the MS-XFEM quality of the solution.

\section*{Acknowledgements}
Fanxiang Xu is sponsored by the Chinese Scholarship Council (CSC). Authors acknowledge Yaolu Liu of TU Delft and all members of the Delft Advanced Reservoir Simulation (DARSim) and ADMIRE research group for fruitful discussions.

\bibliography{main.bib}
\bibliographystyle{ieeetr}

\end{document}